\documentclass[journal=jpclcd,manuscript=article]{achemso}

\usepackage{chemformula} 
\usepackage[T1]{fontenc} 
\usepackage[capitalize]{cleveref}
\crefname{equation}{Eq.}{Eqs.}
\Crefname{equation}{Equation}{Equations}
\usepackage{mathtools}

\author{Zhen Tao}
\email{taozhen@sas.upenn.edu}
\author{Tian Qiu}
\author{Joseph E. Subotnik}
\email{subotnik@sas.upenn.edu}
\affiliation[University of Pennsylvania]
{Department of Chemistry, University of Pennsylvania, Philadelphia, Pennsylvania 19104, USA}

\title
  {Symmetric Post-Transition-State Bifurcation Reactions with Berry Pseudo-Magnetic Fields}

\begin{document}

\begin{tocentry}
\begin{center}
    \includegraphics[width=2in,height=2in]{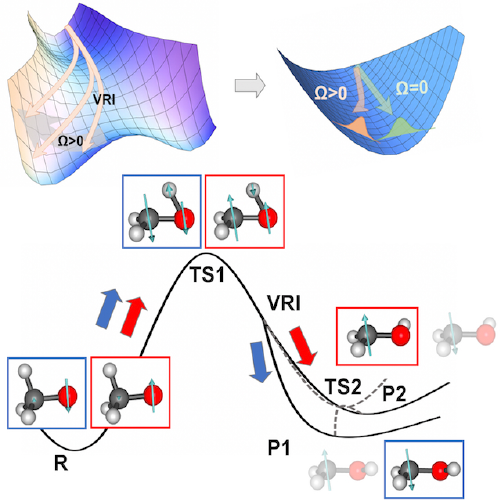}
\end{center}
\end{tocentry}

\begin{abstract}
We investigate how the Berry force (i.e. the pseudo-magnetic force operating on nuclei as induced by electronic degeneracy and spin-orbit coupling (SOC)) might modify a post-transition state bifurcation (PTSB) reaction path and affect product selectivity for situations when multiple products share the same transition state. To estimate the magnitude of this effect, Langevin dynamics are performed on a model system with a valley-ridge inflection (VRI) point in the presence of a magnetic field (that mimics the Berry curvature). We also develop an analytic model for such selectivity that depends on key parameters such as the surface topology, the magnitude of the Berry force, and the nuclear friction. Within this dynamical model, static electronic structure calculations (at the level of generalized Hartree-Fock with spin-orbit coupling (GHF+SOC) theory) suggest that electronic-spin induced Berry force effects may indeed lead to noticeable changes in methoxy radical isomerization.
\end{abstract}

\begin{figure}
  \includegraphics[width=2.9in]{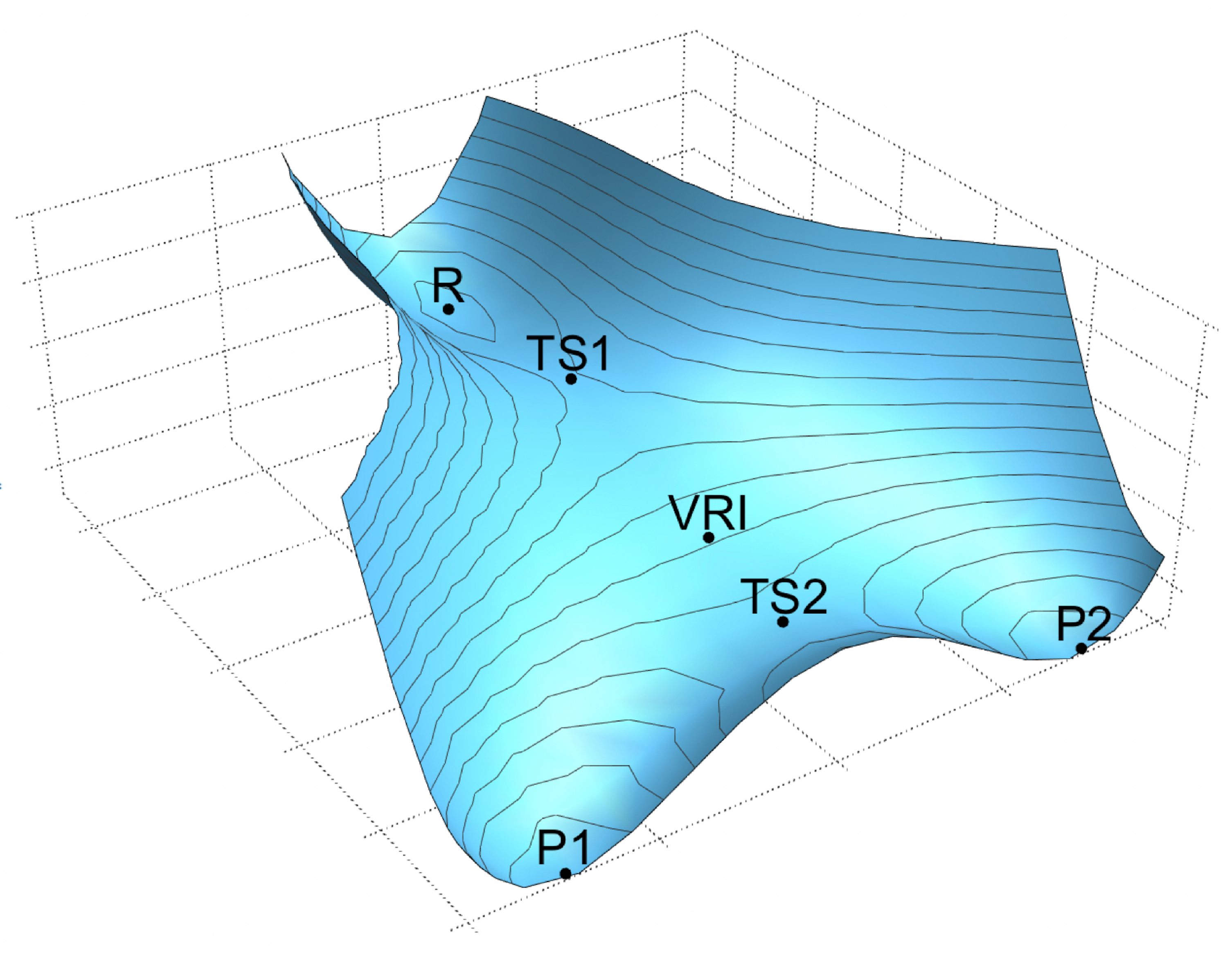}
  \caption{Schematic illustration of a symmetric post-transition state bifurcation reaction. The first transition state (TS1) connects with the reactant well (R) and another transition state (TS2). The latter TS serves as the transition state between the two product wells (P1 and P2). The valley-ridge inflection (VRI) point characterizes the point where the potential energy surface changes from a valley to a ridge, i.e. mathematically where there is a Hessian eigenvector of zero curvature which is orthogonal to the gradient.}
  \label{fgr:vri_potential}
\end{figure}

Post-transition-state bifurcation (PTSB) reactions are an important class of reactions for organic, organometallic, and biosynthetic systems.\cite{Singleton2003,Ess2008,Rehbein2011,Hare2017} With advances in theoretical and experimental tools, an increasing number of PTSB reactions have been reported in a wide range of processes such as rearrangement,\cite{Hansen2011,Hare2018} pericyclic reactions,\cite{Caramella2002,Thomas2008,Wang2009} metal-catalyzed processes,\cite{Hare&Tantillo2017,Li2021} and enzymatic reactions.\cite{Ess2008,Rehbein2011,Major2012,Hare2017,Wang2021} In these reactions, the reaction paths can bifurcate after a single transition state, leading to formation of multiple products\cite{Hong2014,Xue2019} or intermediates that continue to form different products. A typical schematic PTSB potential energy surface is shown in \cref{fgr:vri_potential}. To select the desired reaction pathway and avoid side products in organic synthesis and biosynthesis, it is crucial to understand the factors that affect the product selectivity and be able to accurately predict the product selectivity.

Through previous studies, the community has learned the following about the PTSB governing potential energy surfaces\cite{Ess2008,Chuang2020}. First, there is one reactant well but two transition states (TS1 and TS2 in \cref{fgr:vri_potential}) 
To escape its own well, the reactant must go over TS1, whereupon it can eventually reach two possible products. These two products are separated by TS2, and can be either symmetric or non-symmetric. Second, a valley-ridge inflection (VRI) point can be identified as a point that marks the valley-to-ridge transition on the effective surface and this point is usually identified as the location of the bifurcation (which thus makes such a point of great interest to the community).\cite{Harabuchi2011,Collins2013,Garrido2021} From a mathematical perspective, the VRI point is almost always defined as the point in the configuration space where there is a Hessian eigenvalue with zero curvature and the corresponding is orthogonal to the gradient.\cite{Valtazanos1986} While it is relatively easy to locate the VRI point for the symmetric bifurcation reactions, a VRI point is in general not on the intrinsic reaction coordinate (IRC)\cite{Fukui1981,Maeda2015} for non-symmetric reactions and methods have been developed to locate VRI points for some complicated models of molecules.\cite{Gill1988,Quapp2004,Quapp2011,Schmidt2012} 

Notwithstanding these advances, from a computational and/or theoretical perspective, modeling PTSB reactions remains very difficult; at present, there really is no complete alternative to molecular dynamics.  \cite{Hase2003,Paranjothy2013}
After all, 
PTSB reactions involve multiple products sharing the same transition state, and thus all selectivity is dictated exclusively by dynamical effects (which are inevitably entangled with the topology of the potential energy surfaces).\cite{Carpenter1998,Thomas2008,Rehbein2011,Rehbein2015} Standard statistical reaction rate theories  (transition state theory (TST) and Rice-Ramsperger-Kassel-Marcus theory)\cite{Rice1927,Kassel1928,Marcus1952,Bao2017} simply cannot be applied to predict the product selectivity. Beyond TST, inexpensive alternatives to molecular dynamics are continuously of interest and a few methods that rely only on dynamical and geometric information at key points on the PTSB potential energy surface--such as transition states and products states--have been and continue to be developed aiming to circumvent the high computational cost of propagating trajectories.\cite{Peterson1992,Yang2018,Lee2020,Bharadwaz2021} This active area of research remains a work in progress, and one's ability to predict PTSB selectivity without running trajectories will depend on many details of the potential energy surface as relevant to each specific molecule and/or system. 

One of the major difficulties in predicting PTSB reaction selectivity (especially without running molecular dynamics) is that the product distribution (at least in some cases) has been found to be very sensitive to the local environmental effects; this sensitivity presumably arises due to a lack of a second barrier gating product formation. To date, many different environmental effects have been studied, including solvent effects\cite{Sheppard2009,Carpenter2014,Carpenter2015,Hare2018,Wang2021} and the presence of metal catalysts or enzymes.\cite{Major2012,Hare&Tantillo2017,Li2021,Wang2021} 
Interestingly, Carpenter \cite{Carpenter2014} has predicted that applying a chiral external electric field can have large microscopic effects on the product selection, but the net effect will be averaged away due to the free rotation of the molecules. To our knowledge, the effect of a magnetic field has not been investigated for this kind of system, likely for the same reason. That being said,  if one could construct a large magnetic field in a molecular frame, there is no reason not to presume that such a magnetic field could  greatly influence PTSB dynamics. 

At this point, it would seem natural to turn our attention to weak nonadiabatic effects.  Nonadiabatic effects are those effects that arise when there are nearby excited states that must be accounted for, so that molecular dynamics no longer run on a single Born-Oppenheimer surface.
Moreover, it is known that (nonadiabatic) vibronic interactions with nearby excited states of specific symmetries can lead to bifurcation through the second-order Jahn-Teller distortions.\cite{Taketsugu1994,Harabuchi2011}  Thus, one must wonder,  might nonadiabatic effects affect PTSB through a VRI point? In light of the previous paragraph, this question is particularly apt because, when molecules visit regions of configuration space with degenerate electronic states and spin-orbit coupling, an internal magnetic field (a so-called ``Berry'' force)\cite{Truhlar1979,Berry1984,Subotnik2019,Bian2021} arises in the molecular frame. For a given state $j$, nuclei of mass $m$ and momentum $\mathbf{P}$, the Berry force is of the form.\cite{foot}
\begin{equation}
    \label{eqn:berry_force}
    \mathbf{F}^{\text{Berry}}_{j}  = 2\hbar \operatorname{Im}\sum_{k\ne j}\Big[\mathbf{d}_{jk}\Big(\frac{\mathbf{P}}{m}\cdot\mathbf{d}_{kj}\Big)\Big],
\end{equation}
where $\mathbf{d}_{jk}=\langle\psi_j|\frac{\partial}{\partial \mathbf{R}}\psi_k\rangle$ is the derivative coupling vector between the electronic wavefunction $\psi$ of state $j$ and state $k$ and $\mathbf{R}$ is a set of nuclear coordinates. This Berry force arises because the phase of the ground-state energy changes as a function of position in the presence of spin-orbit couplings and/or magnetic fields, where the Hamiltonian becomes complex-valued. And so the question of whether (nonadiabatic)  magnetic field effects can affect PTSB becomes quite relevant. The suggestion\cite{Wu2020,Wu2021,Bian2021} has been made recently that Berry forces and nonadiabatic effects may be responsible for the chiral induced spin selectivity (CISS)\cite{Waldeck2012,Naaman2019,Evers2022} phenomenon, whereby enantioselective reactions can be achieved using spin-polarized electrons. \cite{Bloom2020}

In order to better understand the consequences of  a hypothetical magnetic field effect on a PTSB reaction, consider \cref{eqn:berry_force} for the case of a molecule with an odd number of electrons.  In such a case, it can be shown that if the net spin of the molecule is up, then the Berry force operating on the molecule will be exactly equal and opposite to the Berry force that the molecule would experience if the net spin were down.\cite{Wu2020}  Thus, the internal pseudo-magnetic Berry force that operates on the nuclear degrees of freedom reflects the spin state of the electronic degrees of freedom, and so if the Berry force can indeed lead to PTSB product selectivity, that selectivity would reflect the electronic state of the molecule.  This interesting state of affairs would constitute a new approach to selectivity.

With this background in mind, the goals of the present letter are as follows. First, we will go beyond previous studies of Berry force dynamics on flat potentials\cite{Miao2019,Bian2021} and, using a Langevin equation, we will consider if and how Berry forces can guide PTSB selectivity in the presence of damping and random force effects. Second, whereas previous work has centered on purely model Hamiltonians, our goal will be to work with a reasonable VRI model and estimate the 
smallest  Berry force  that would be needed to meaningfully affect the product selectivity in a condensed phase reaction. 
Third, given that molecular dynamics are expensive and one would like to extract as much information as possible, we will derive an analytic model for estimating such a minimum Berry force in terms of the topology of the potential energy surface. Fourth and finally, we will work with a standard reaction known to exhibit a VRI (methoxy radical isomerization),\cite{Colwell1984,Colwell1988,Gill1988,Taketsugu1996,Yanai1997,Kumeda2000,Lasorne2003,Ess2008,Harabuchi2011,Maeda2015,Chuang2020} whereby we can estimate whether such a minimum Berry force is realizable for a real system. These goals will provides a stepping-stone for the future {\em ab initio} molecular dynamics studies of the realistic molecular systems.

To study the potential effect of a Berry force on the selectivity of bifurcation reactions, Langevin dynamics were performed on a two-dimensional (2D) model potential $V(x,y)$ and a uniform magnetic field was applied in the positive z direction to represent the pseudo-magnetic  Berry force. 
\begin{equation}
    \label{eqn:langevin}
    \begin{split}
        \ddot{x} &= - \gamma\dot{x}   -\frac{\nabla_{x}V(x,y)}{m} + \frac{\Omega}{m} \dot{y} 
         + \frac{\eta_{\textrm{x}}(t)}{m}\\
        \ddot{y} &= - \gamma\dot{y} -\frac{\nabla_{y}V(x,y)}{m} - \frac{\Omega}{m} \dot{x} 
         + \frac{\eta_{\textrm{y}}(t)}{m}
    \end{split}
\end{equation}

Following standard notation, in \cref{eqn:langevin}, $\gamma$ is the friction coefficient and represents the system-bath coupling. A Gaussian white noise $\eta_{\alpha}(t)$ with a mean of zero $\langle \eta_{\alpha}(t)\rangle = 0$ is used to model the thermal fluctuations in the bath, where the indices $\alpha$, $\beta$ represent  Cartesian directions $\{x,y,z\}$.  The noise is  delta-correlated  (which assumes that the relaxation time of the bath is much faster than the timescales of the system) and satisfies the fluctuation-dissipation theorem $\langle \eta_\alpha(t)\eta_\beta(t')\rangle =2\gamma k_{\textrm{B}}T m\delta_{\alpha\beta}\delta(t-t')$.  Here, the constant $k$ is the Boltzman constant and $T$ is the temperature. The quantity $\Omega$ is the off-diagonal element of the 2-by-2 anti-symmetric Berry curvature matrix in this 2D problem, which is defined as 
\begin{equation}
    \label{eqn:berry_curvature}
    \mathbf{\Omega}_{j} =i\hbar\nabla \times \mathbf{d}_{jj}
\end{equation}\\
 In this framework,  the Berry force operating on surface $j$ is $\mathbf{F}^{\text{Berry}}_{j}  = \mathbf{\Omega}_{j}\mathbf{v}$. (Note that, if one differentiates both the ket and the bra of $\mathbf{d}_{jj}=\langle\psi_j|\frac{\partial}{\partial \mathbf{R}}\psi_j\rangle$ in \cref{eqn:berry_curvature}, and then inserts a complete sum of states, one recovers \cref{eqn:berry_force}.)

\begin{figure*}
  \includegraphics[width=5.8in]{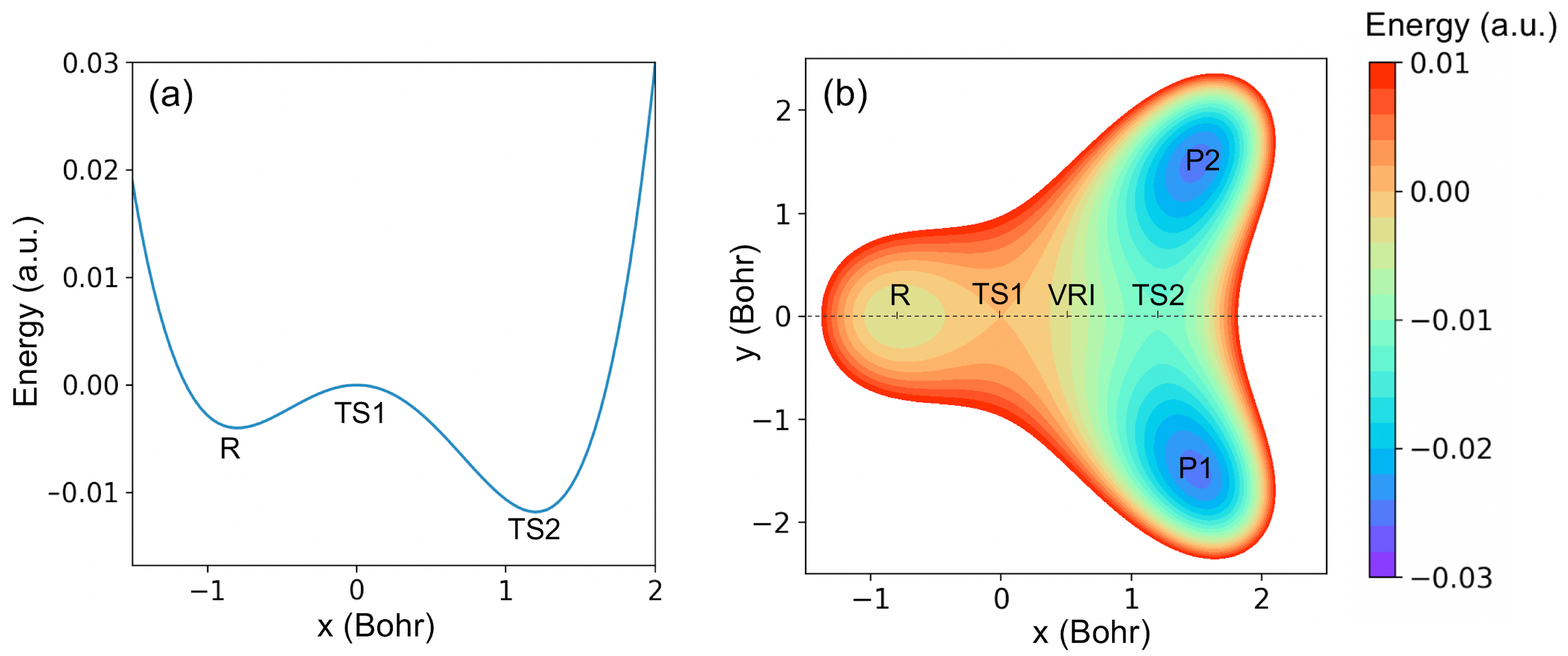}
  \caption{(a) The 1-D slice of the asymmetric potential along the $y=0$ axis and (b) the 2-D contour plot of the symmetric PTSB model potential. The model form is given in Eq.\eqref{eqn:model}, with parameters $A = \frac{1}{4}$, $B=\frac{2}{3}$, $C=12$, $V^{\ddagger}=\frac{3}{64000}$, $x_{\textrm{s}}=0.2$. The parameters $D$, $F$, $G$ are determined to be 0.01534, 0.00310, 0.00006, respectively when setting the product wells (P1 and P2) to be at the coordinates of $(1.5,\pm 1.5)$ with energy $-0.025$ Hartree. The VRI point is set to be at the coordinate of $(0.5,0.0)$ for this plot.}
  \label{fgr:potential}
\end{figure*}

The specific form of the symmetric PTSB model potential $V(x,y)$ is given in \cref{eqn:model} and has been adapted 
from a previously studied potential\cite{Garrido2021} in two ways. The first term in \cref{eqn:model} has been changed so that the reactant well and TS2 no longer have the same energy. (See \cref{fgr:potential}(a)). The last term has also been slightly modified to ensure the product wells are bound for the parameters of interest. \cref{fgr:potential}(b) gives a representative contour plot of the model potential.   
\begin{equation}
    \label{eqn:model}
    \begin{split}
        V(x,y) =
        &\frac{V^{\ddagger}}{x^4_{\textrm{s}}}x^2(Ax^2-Bxx_{\textrm{s}}-Cx^2_{\textrm{s}})\\
        &+Dy^2(x_{\textrm{i}}-x)+y^4(F+Gx^4)
    \end{split}
\end{equation}
For this study, the parameters $A$,$B$ and $C$ are set to be $\frac{1}{4}$, $\frac{2}{3}$, and $12$, respectively, for which  reactant well R, the TS1, and the TS2 are located at points: $(-4x_{\textrm{s}},0.0)$ with energy -$\frac{256}{3}V^{\ddagger}$, $(0.0,0.0)$ with energy 0.0, and $(6x_{\textrm{s}},0.0)$ with energy $-252V^{\ddagger}$. The parameter $x_{\textrm{s}}$ sets the separation between the reactant well and the TS2, and is chosen to be 0.2 Bohr. The parameter $V^{\ddagger}$ sets the depth of the reactant well and the energy difference between TS1 and TS2, which is chosen as $\frac{3}{64000}$ to make the reaction barrier TS1 equal to 0.004 Hartree ($\sim4k_{\textrm{B}}T$ for T = 300K) (and thus amenable to direct computation).  All the units above and below are given in atomic units (a.u.) unless stated otherwise.  In the second and third terms, the parameter $x_{\textrm{i}}$ can be used to set the position of the VRI point at $(x_{\textrm{i}},0.0)$.  Similar to the original model in Ref. \cite{Garrido2021}, the last three parameters, $D$, $F$, $G$ were determined by setting the coordinates of the product wells to be at $(x_p,\pm y_p)$ and their energies to be $-0.025$ Hartree. In this work, with $x_p = y_p = 1.5 $, the parameters $D$, $F$, $G$ can be determined based on the choice of $x_{\textrm{i}}$ (see the section S1 of the Supporting Information).  

Langevin dynamics were initialized in the reactant well using the symmetric PTSB potential above.  We propagated 20000 trajectories ($N_{\textrm{total}}$) and calculate the number of trajectories ending in P2 well ($N_{\textrm{P2}}$) versus P1 well ($N_{\textrm{P1}}$). The product selectivity was calculated as 
\begin{equation}
    S(\Omega,x_{\textrm{i}},\gamma,T,m) = \frac{N_{\textrm{P1}}-N_{\textrm{P2}}}{N_{\textrm{total}}}\cdot100\%
\end{equation}
 The particle mass was fixed at 2000 a.u. (the mass of an electron is 1 a.u.) and temperature was fixed at 300K for all the trajectories, so the product selectivity can be safely written in shorthanded as $S(\Omega,x_{\textrm{i}},\gamma)$. Each trajectory was propagated using a Verlet-type propagator with a magnetic field \cite{Spreiter1999,Jensen2013} using a time step of 0.242 fs and the details for this propagator are given in the section S2.1 in the Supporting Information. The barrier between P1 and P2 is $\sim 14k_{\textrm{B}}T$, which delays thermal redistribution of the products (to nanoseconds or longer).   Thus, to save computational costs, we terminated the simulation of each trajectory after the particle began its downhill march to P1 or P2 and the new potential energy was $3k_{\textrm{B}}T$ below the energy of TS2; in practice, for all damping parameters, we find kinetic energy relaxation immediately after a trajectory enters a product well (and any P1-P2 recrossing should be thermally activated rather than ``hot  immediate recrossings'').  Convergence tests for the number of trajectories and the propagation time step are provided in the section S2.2 of the Supporting Information.

\begin{figure*}
  \includegraphics[width=5.8in]{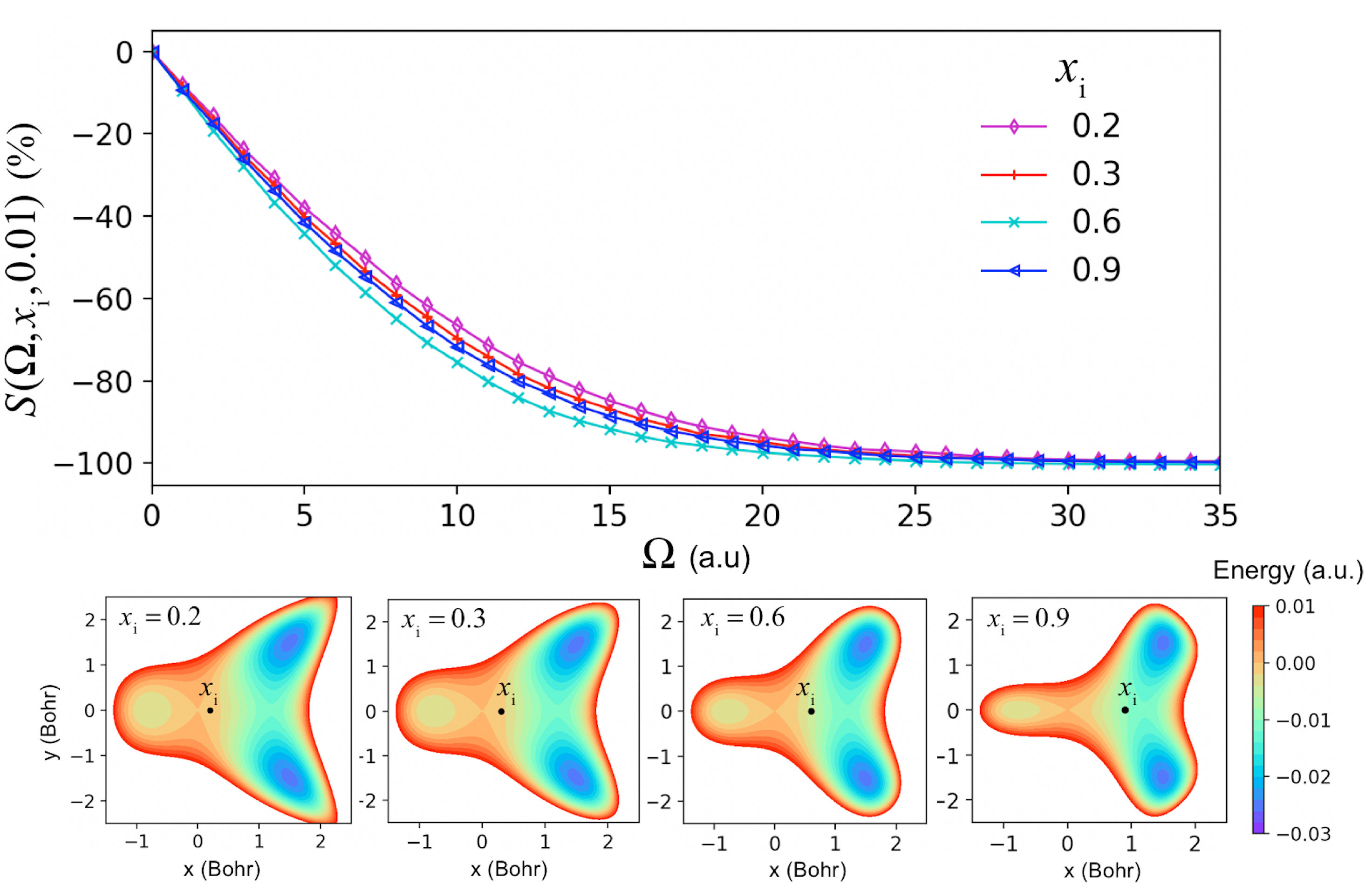}
  \caption{Top panel: impact of the strength of the effective magnetic field on the product selectivity $S(\Omega,x_{\textrm{i}},\gamma)$ with different choices of VRI locations $x_{\textrm{i}}$. Bottom four panels: effect of the VRI locations on the model potential. All of the VRI points are selected to be on the $y=0$ axis, so the different VRI positions are defined by the x coordinates ($x_{\textrm{i}}$) of the VRI point. As the VRI position moves away from TS1 or the value of $x_{\textrm{i}}$ increases, the impact of the effective magnetic field on the product selectivity first increases and then decreases. As the VRI point is set closer to the TS2, the width of the TS1 along the y-axis becomes narrower and the slope of the valley before the VRI point becomes steeper. }
  \label{fgr:selectivity_B}
\end{figure*}

In the top panel of \cref{fgr:selectivity_B}, we plot the results for product selectivity with different values of the Berry curvature element. We find that, for these parameters, in order for the magnetic field to have any effect, the magnitude has to be at least on the order of $\pm10^{-1}$ a.u. (equivalent to $\sim 10^4$ Tesla for an uniform external magnetic field). To get a sizable effect on the selectivity, $\Omega$ needs to be still one order of magnitude larger. With the value of the $\Omega$ is even larger, on the order of ten (i.e. $\sim 10^6$ Tesla for an uniform external magnetic field), the selectivity is completely determined dominantly by the sign of the magnetic field. Although this magnitude might seem extremely large for an external magnetic field,  
as will shown below, internal (Berry) pseudo-magnetic fields can be much larger than external fields. For instance, 
according to (\cref{eqn:berry_force})
(which expresses the Berry curvature in term of derivative couplings), the Berry curvature can explode in the vicinity of an avoided crossing regions or conical intersection (albeit in a localized area of configuration space).

In \cref{fgr:selectivity_B},  we  also investigate the effect of the VRI location on product selectivity by propagating dynamics with different $x_{\textrm{i}}$ values (as chosen between the positions of TS1 and TS2 in \cref{fgr:potential}(a)).  The relevant PESs are shown in the bottom four panels in \cref{fgr:selectivity_B}. In general, the closer the VRI point to the TS2 (larger $x_{\textrm{i}}$), the narrower the reactant well and the TS1 barrier width, and the steeper the slope of the valley connecting TS1 and the VRI point. In \cref{fgr:selectivity_B}, we find that when we start from a small $x_{\textrm{i}} = 0.2$ and increase $x_{\textrm{i}}$ (so that the VRI moves away from TS1), product selectivity reaches a maximum when $x_{\textrm{i}} = (0.6,0.0)$. The selectivity then decreases as $x_{\textrm{i}}$ further increases and the VRI location moves more toward TS2. As we will now explain, this observed trend arises from two competing factors: the width of the TS1 barrier along the $y$-axis and the steepness of the slope of the valley along the $x$-axis. 

To explain these observations from the simulations, we have generated a simple model whereby we substitute for the symmetric PTSB model potential in \cref{eqn:model} a  surface of the form:
\begin{equation}
    \label{eqn:simplified_model}
    \begin{split}
        V^{\textrm{s}}(x,y) = -\sigma_{\textrm{x}}x+\frac{k_{\textrm{y}}y^2}{2}
    \end{split}
\end{equation}
 with a negative linear slope $-\sigma_{\textrm{x}}$ along the x axis and a positive quadratic slope $k_{\textrm{y}}$ in the y direction. The linear slope along the x direction $-\sigma_{\textrm{x}}$ can be estimated from the energy drop between TS1 and the VRI point over the distance between them. As TS1 is fixed at coordinate $(0.0,0.0)$ with energy $0.0$ in \cref{eqn:model}, the linear slope $\sigma_{\textrm{x}}(x_{\textrm{i}}) = -\frac{V(x_{\textrm{i}},0)}{x_{\textrm{i}}}$ is a simple function of  $x_{\textrm{i}}$. Unfortunately, for the potential in \cref{eqn:model},  $k_{\textrm{y}}$ is not constant (as a function of $x$) and must be approximated. To that end, we discretize the distance between  TS1 and the VRI point into 400 grid points and we choose $k_{\textrm{y}}$ as that second derivative that one calculates at point number 351; thereafter, $k_{\textrm{y}}$ is also a simple function of $x_{\textrm{i}}$  that can be calculated numerically. More details regarding the choice of $k_{\textrm{y}}$ and the specific values of $\sigma_{\textrm{x}}$ and $k_{\textrm{y}}$ are given in the section S4 in the Supporting Information. Using the simplified model potential in \cref{eqn:simplified_model}, Langevin dynamics--with and without an effective magnetic field--were investigated analytically in the overdamped limit. Results are provided in the section S3 of the Supporting Information. Our main results in the overdamped limit are that the average position along the y direction $\langle y(t)\rangle_{t\rightarrow\infty}$ without and with the effective magnetic field are found to be $0.0$ and $- \frac{\sigma_{\textrm{x}}\Omega}{m\gamma k_{\textrm{y}}}$, respectively. The variance $\langle( y(t) - \langle y(t)\rangle)^2\rangle_{t\rightarrow\infty}$ of the distribution of the y values is $\frac{ k_{\textrm{B}}T}{k_{\textrm{y}}}$, whether there is an effective magnetic field or not. 
 
Using the mean and variance of the $y$-distribution, we can estimate the selectivity for the simplified model above.  If we assume the distributions (with $f(x)$ and without a field $g(x)$) are Gaussian, we can calculate their (unitless)  Hellinger distance $H$:
 \begin{equation}
\label{eqn:hsquare}
 H^2(f,g) = \frac{1}{2} \int\Big(\sqrt{f(x)}-\sqrt{ g(x)}\Big)^2 dx
 \end{equation}
Note that the Hellinger distance is bound within the range between zero and one, as one would expect for the product selecitivity.  Moreover, a Hellinger distance of zero 
implies that the field has no impact on the product selectivity, which also corresponds to a zero selectivity. In the other limit, a Hellinger distance of one means the presence of the effective magnetic field completely changes the position distribution, which can be understood as a 100\% of selectivity. When applied to the overdamped model above, if one calculates the Hellinger distance between the two distributions, the result is: 
\begin{equation}
    \label{eqn:helinger}
    \begin{split}
         H(\Omega,x_{\textrm{i}}, \gamma, T, m) =\Big(1-e^{-\frac{(\sigma_{\textrm{x}}(x_{\textrm{i}}))^2\Omega^2}{8 k_{\textrm{B}}Tk_{\textrm{y}}(x_{\textrm{i}})m^2\gamma^2}}\Big)^\frac{1}{2}\cdot100\%
    \end{split}
\end{equation}
The particle mass is fixed at 2000 a.u. and temperature is fixed at 300K for all the trajectories, so the product selectivity is shorthanded as $H(\Omega,x_{\textrm{i}},\gamma)$ for the discussions below.

\begin{figure}[H]
  \includegraphics[width=2.9in]{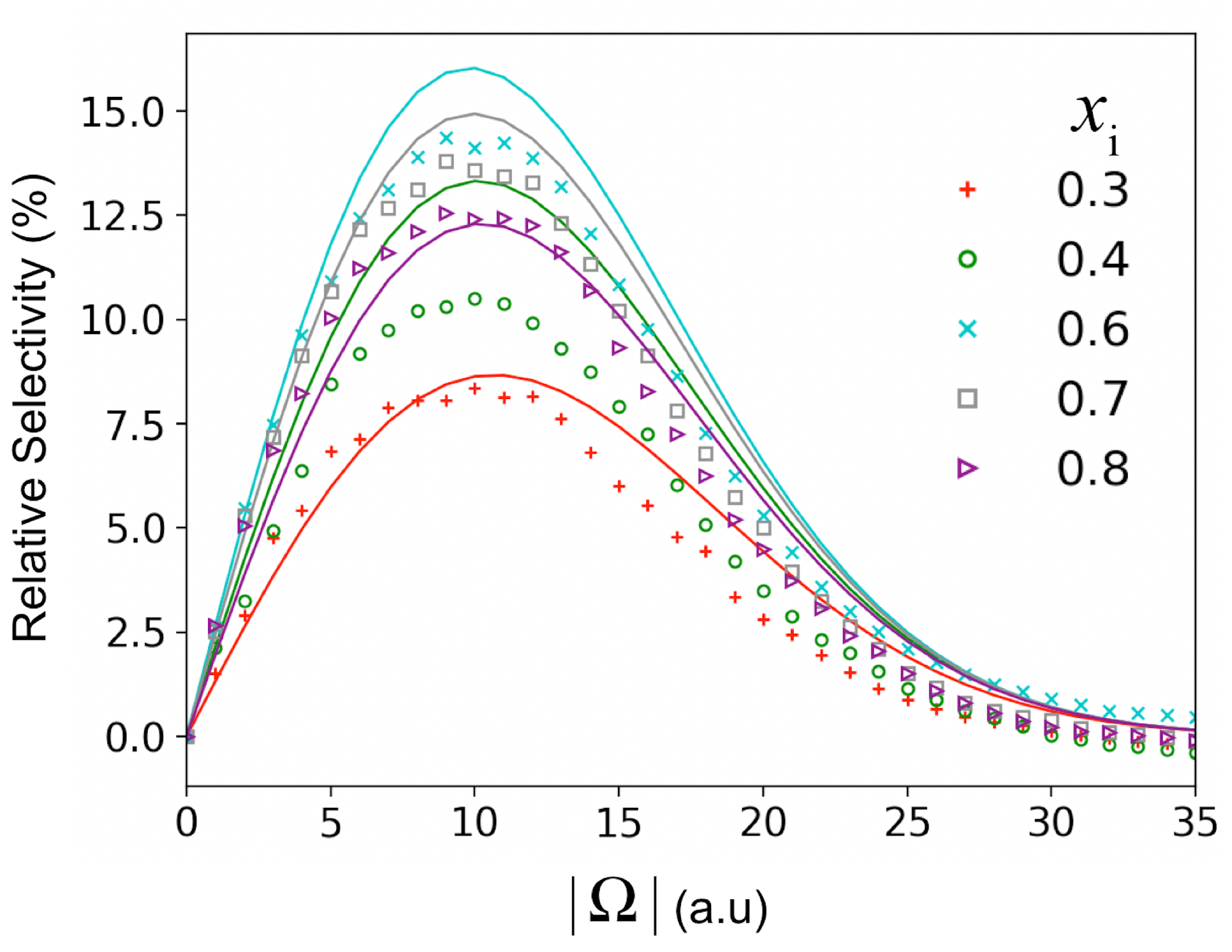}
  \caption{
  A plot of the product selectivity relative to $H(\Omega,0.2,0.01)$ as a function of the magnitude of the Berry curvature element computed and for different VRI locations ($x_{\textrm{i}}$) according to (i) the simulations (no-lines) and (ii) the analytic solutions of a simplified model (lines).} Both the simulation and the analytic theory show the product selectivity first increases as the VRI point moves closer from TS1 to TS2, and then decreases.
  \label{fgr:relative_vri}
\end{figure}

In \cref{fgr:relative_vri},  we compare the the relative selectivity of our simulations ($S(\Omega,x_{\textrm{i}},0.01) -  H(\Omega,0.2,0.01)$) versus the relative Hellinger distances of the analytic model ($H(\Omega,x_{\textrm{i}},0.01,) - H(\Omega,0.2,0.01)$)  as a function of the magnetic field (which mimics the Berry curvature). We find that, with $\gamma=0.01$, the product selectivity with different values of the Berry curvature element estimated by \cref{eqn:helinger} gives a trend similar to what is found in the simulation results.  Moreover, we find an empirical relationship between the VRI locations and the product selectivity, whereby the selectivity first increases as the VRI point moves closer from TS1 towards TS2 and then decreases. This maximum can be understood as arising from two competing factors: the width of the TS1  ${(k_{\textrm{y}})}$ versus the slope between TS1 and the VRI point $(\sigma_{\textrm{x}})$, as shown from the ratio of $\frac{\sigma_{\textrm{x}}^2}{k_{\textrm{y}}}$ in the exponent in \cref{eqn:helinger}. In addition, according to both approaches, we find a maximum in the impact of the VRI location on the product selectivity for an effective magnetic field around 11 a.u.;  for larger magnetic fields, the magnetic field dominates such that there is no longer a relative field effect.
Note that an underlying assumption of the analytic theory is that the product selectivity is solely determined by the particle's position when reading  the vicinity of the VRI point. Of course, such an assumption has limitations, as it is possible for a particle with a position to the right of the VRI point to end up in the opposite product well due to its momentum directions pointing towards the opposite product well. This effect is termed as "dynamic matching"\cite{Carpenter1998} and might explain the quantitative differences observed here between the simulation results and the analytic theory.  Simplification of the potential energy surface can also contribute to these quantitative differences. 

\begin{figure}
  \includegraphics[width=2.9in]{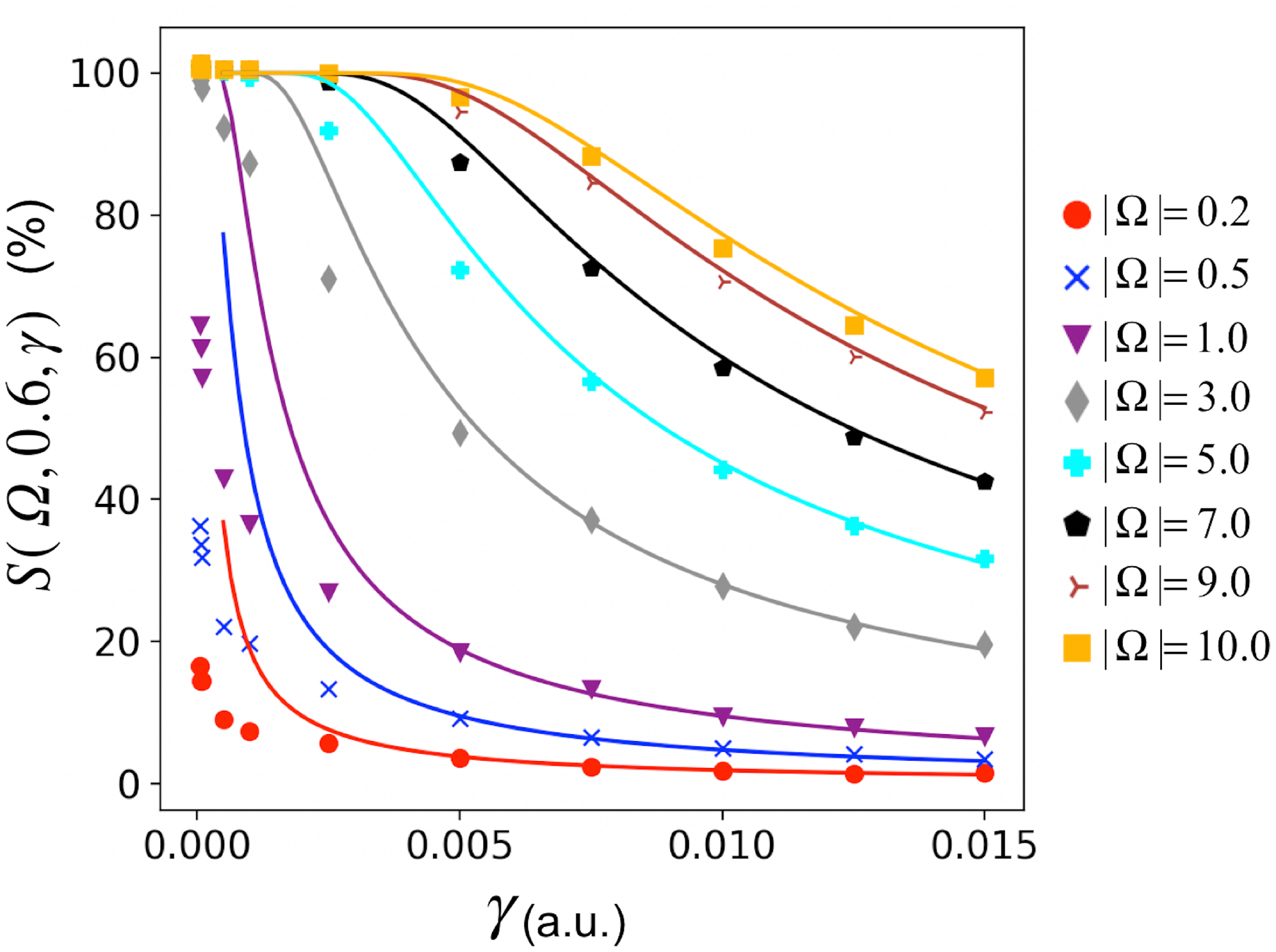}
  \caption{Effect of the friction coefficient on the product selectivity according to simulations ($S(\Omega,0.6,\gamma)$, no-lines) and  analytic theory based on a simplified model ($H(\Omega,0.6,\gamma)$, lines). The simulation data is captured well  by our model for large $\gamma$ ($\gamma\ge 0.005$).  Note that, for small $\gamma$, the selectivity predicted by simulations can be large (even for small $\Omega$).  However, in such a limit outside the overdamped regime, the analytic result in \cref{eqn:helinger} is not valid.  }
  \label{fgr:fric}
\end{figure}

Lastly, in \cref{fgr:fric}, we investigate the impact of the friction coefficient $\gamma$ on the product selectivity according to our simulations and the analytic theory. For a variety of different strengths of the effective magnetic fields, the analytic results and the simulations agree for large $\gamma$ ($\gamma\ge 0.005$); as the friction coefficient increases, the effect of the effective magnetic field on the product selectivity decreases. For small values of $\gamma$, according to simulations, one can recover a reasonable selectivity even for small magnetic field.  However, at small values of $\gamma$, the analytic theory overestimates the selectivity and goes up to $100\%$ selectivity too quickly, because the overdamped formalism leading to \cref{eqn:helinger} no longer holds. Future work to extend \cref{eqn:helinger} beyond the overdamped regime will be helpful.

\begin{figure*}
    \centering
    \includegraphics[width=5.8in]{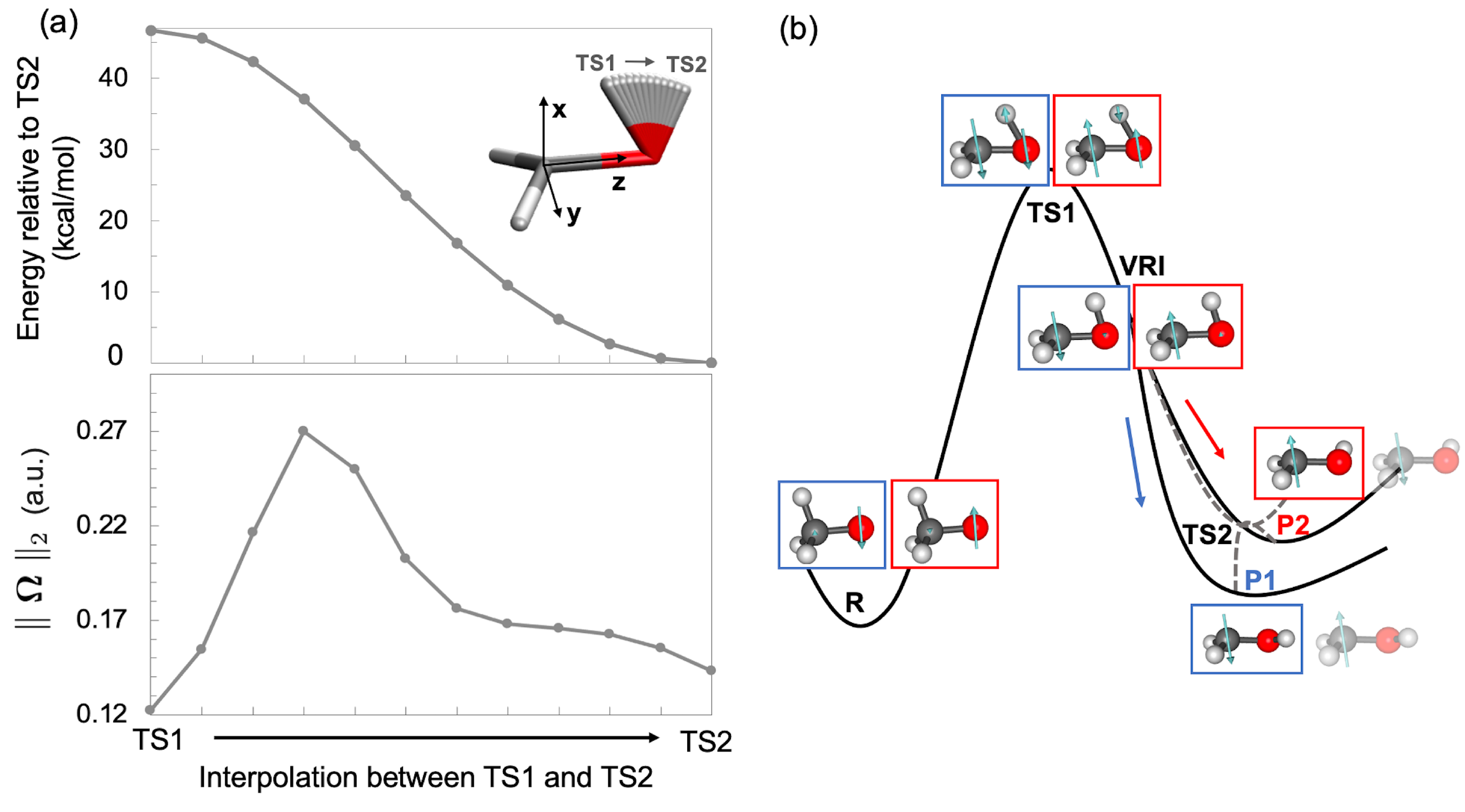}
    \caption{(a) Top panel: GHF+SOC/6-31G(d,p) energies along the TS1-to-TS2 path of methoxy radical relative to TS2 energy. The changes in the methoxy radical geometries along the TS1-to-TS2 path are shown in the top right section of the top panel.  Bottom panel: Magnitudes of Berry curvature along the TS1-to-TS2 path. The magnitude of Berry curvature at a given geometry is defined by the induced 2-norm $\parallel \mathbf{\Omega} \parallel_{2}$, which corresponds to the largest eigenvalue of $\sqrt{\mathbf{\Omega}^{T}\mathbf{\Omega}}$. (b) A schematic plot showing how the direction of the spin magnetic moment\cite{Hoyos2014} can bias  product formation. The directions of atom-based spin magnetic moments are shown by arrows in cyan. The states surrounded by blue and red boxes are related to each other by  time reversal symmetry.}
    \label{fgr:BC}
\end{figure*}

Finally, before concluding, we would like to apply the theory above to a realistic molecular system, the methoxy radical isomerization (see Fig. \ref{fgr:BC}), which is a prototypical example of a symmetric PTSB reaction.\cite{Colwell1984,Colwell1988,Gill1988,Taketsugu1996,Yanai1997,Kumeda2000,Lasorne2003,Ess2008,Harabuchi2011,Maeda2015,Chuang2020}  Is the necessary field strength ($\parallel \mathbf{\Omega} \parallel = 0.2$ a.u.) required in Fig. \ref{fgr:fric} achievable and/or realistic ?
To that end, we have calculated the Berry curvature\cite{Culpitt2021,Culpitt2022} near the TS1-to-TS2 reaction path using generalized Hartree-Fock with spin-orbit coupling interaction (GHF+SOC) \cite{Pople1977,Fukutome1981,Goings2015} and complex orbitals; the spin-orbit coupling operator used is the one-electron component of the Breit-Pauli form.\cite{Abegg1975} The TS1-to-TS2 path was estimated by performing linear interpolation between the internal coordinates of the optimized TS1 and TS2 structures as obtained from previously generated  unrestricted HF calculations \cite{Taketsugu1996} with a 6-31G(d,p) basis set. This computation was performed in a local branch of Q-Chem 6.0.\cite{Epifanovsky2021} Additional details about the method and the implementation can be found in section S5 in the Supporting Information. 

The GHF plus SOC energy and the magnitudes of Berry curvatures are plotted along the TS1-to-TS2 path in \cref{fgr:BC}(a). The VRI point is close to the TS1-to-TS2 path \cite{Taketsugu1996} and the magnitude of the Berry curvature at the VRI point is calculated to be 0.17 a.u.. Note that the Berry force does not change much in magnitude as one traverses the TS1-to-TS2 path for methoxy radical. This relative invariance is important as it justifies the use of the model in 
\cref{eqn:langevin}, where assumed that the magnetic field was constant. Interestingly, even though the Berry force can be expanded formally as a sum over derivative couplings (see \cref{eqn:berry_force}) which often depend sensitively on geometry and diverge near a conical intersection, for this system with an extra unpaired electron, there is a background level of Berry curvature that is relatively constant.

 Lastly, the model described above was two-dimensional whereas the configuration space for the methoxy radical has formally 9 dimensions ($ 3N_{\textrm{atoms}} -6)$. To decipher just how accurate a two dimensional model can be,  we have performed the following calculations. First, we calculated the Berry force along the TS1-to-TS2 path for methoxy radical using a velocity vector pointing in the direction corresponding to the eigenvector with a negative eigenvalue at TS1. Second, we then computed the dot products of the normalized Berry force vector ($\mathbf{F}^{\rm Berry}$) and the unit vector ($\mathbf{v}_{\textrm{TS2}\rightarrow \textrm{P1}}$) pointing from the TS2 to the P1 structure (\cref{fgr:BC}(b)), the latter of which was approximated by the eigenvector corresponding to a negative eigenvalue at TS2. The two eigenvectors used are shown in Fig.S1 in the Supporting Information. The results of such dot products are shown in 
\cref{table:dotp}. The fact that these values are close to 1 would seem to justify the 2D model above. 

We conclude that, for this reaction, the spin state of the molecule should meaningfully bias the reaction towards one product. For instance, for the spin state denoted in the blue box in \cref{fgr:BC}(b), the Berry force will push the molecule towards the P1 product. Vice versa, for the time-reversed spin-state solution in the red box in \cref{fgr:BC}(b) (with spin magnetic moments pointing in the opposite direction), the corresponding equal-and-opposite Berry force will guide the molecule towards P2 formation.  

\begin{table} [H]
    \centering
        \caption{Overlaps between the Berry Force Direction and the TS2-to-P1 Direction \label{table:dotp}}
        \begin{tabular} 
        {p{2cm} p{4cm}}
        \hline \hline
        \hfil Geometry  &  \hfil $\mathbf{F}^{\rm Berry}\cdot\mathbf{v}_{\textrm{TS2}\rightarrow \textrm{P1}}$  \\ 
        \hline
        \hfil TS1 & \hfil  0.56   \\ 
        \hfil Step 1 & \hfil  0.72   \\ 
        \hfil Step 2 & \hfil 0.83  \\
        \hfil Step 3 & \hfil 0.86 \\ 
        \hfil Step 4 & \hfil 0.85  \\ 
        \hfil Step 5 & \hfil 0.84  \\ 
        \hfil Step 6 & \hfil  0.82 \\ 
        \hfil Step 7 & \hfil  0.80  \\ 
        \hfil Step 8 & \hfil 0.80 \\ 
        \hfil Step 9 & \hfil 0.79  \\ 
        \hfil Step 10 & \hfil 0.79  \\ 
        \hfil TS2 & \hfil 0.78  \\ 
        \hfil VRI & \hfil 0.84  \\ 
        \hline\hline
        \end{tabular}
    \end{table}
In conclusion, in this letter, we have used Langevin dynamics and analytic models to study symmetric PTSB reactions in an effective magnetic field. We found that for a particle with an effective mass of 2000 a.u. sitting in a reasonable potential with a VRI point in the limit of an overdamped reaction coordinate, the magnitude of the magnetic field or the Berry curvature needs to be at least on the order of $\pm10^{-1}$ a.u. to have an effect on the product selectivity. While such a magnetic field cannot be realized as an external magnetic field,  using the methoxy radical isomerization as an example, we have demonstrated that this field strength may be achievable for an internal effective magnetic field. More generally, we can expect that internal Berry forces   can  be far larger than external fields (though in principle, one would like to explore the effects of both internal and external fields\cite{Wibowo2021} in the future).  

Several caveats should now be offered with regards to our conclusions.  

First,  our entire treatment here has been based on the premise that all dynamics are overdamped, and it is likely that magnetic field effects will be stronger with weaker solvent effects so that we operate in the critically or even underdamped regime. More generally,  a Langevin treatment cannot capture many crucial details that  arise if we propagate an explicit (rather than implicit) solvent. Second, as the barrier of the TS2 could be low in actual chemical systems, unlike the assumption made in the simulation, the products could redistribute after reaching the product minima, which adds in another complexity.  This effect will certainly depend on the nature of the molecule under investigation. Third and finally, many reactions in reality display non-symmetric bifurcations, which can favor one product over the other and which must be taken into account when analyzing Berry force effects.

Notwithstanding all of these simplifications, the present simulations and analytic theory provide a prediction for the product selectivity as a function of key factors such as the effective mass, friction coefficient, the strength of the effective magnetic field, and the shape of the potential energy surface. Future {\em ab initio} work will be needed to explore realistic surfaces and reactions, extract parameters, and assess whether or not (at the end of the day) electronic spin effects can indeed dictate the direction of a chemical reaction with nuclear degrees of freedom. Future work should also explore how the conclusions presented here may be modified near a metal surface, where there are many more electronic states, there can be large spin-orbit coupling, and the CISS effect is known to be large.\cite{Naaman2019,Zollner2020,Teh2021}

\begin{acknowledgement}
We thank Barry Carpenter, Hung-Hsuan Teh, Yanze Wu, Xuezhi Bian, Clàudia Climent, Xinchun Wu, and Robert Knowles for several interesting conversations.
This work was supported by the U.S. Air Force Office of
Scientific Research (USAFOSR) under Grants No. FA9550-18-1-0497 and No. FA9550-18-1-0420.
\end{acknowledgement}
\begin{suppinfo}
The Supporting Information is available free of charge.
\begin{description}
  \item The details of the parameters for the symmetric PTSB model potential; equations for the Verlet-type propagator with a magnetic field; convergence tests of time step and number of trajectories; analytic solutions of the simplified model with and without the effective magnetic field; determination of the parameters for the simplified model potential; details for implementation and computation of Berry curvatures with GHF+SOC.
\end{description}
\end{suppinfo}

\bibliography{main}

\begin{mcitethebibliography}{74}
\bibitem[foo()]{foot}
The meaning of Berry curvature for a multi-electronic molecule with an even
  number of electrons is usually more complicated. For a review, see Ref. 46.
  \relax
\mciteBstWouldAddEndPunctfalse
\mciteSetBstMidEndSepPunct{\mcitedefaultmidpunct}
{}{\mcitedefaultseppunct}\relax
\EndOfBibitem
\end{mcitethebibliography}

\end{document}


\newpage
\tableofcontents

\newpage

\section{Parameters for the Model Potential}
The expression for the symmetric post-transition-state bifurcation (PTSB) model used in the main letter is also given in \cref{eqn:model} and is adapted from Ref.\cite{Garrido2021} with slight modifications.
\begin{equation}
    \label{eqn:model}
    \begin{split}
        V(x,y) =
        &\frac{V^{\ddagger}}{x^4_{\textrm{s}}}x^2(Ax^2-Bxx_{\textrm{s}}-Cx^2_{\textrm{s}})+Dy^2(x_{\textrm{i}}-x)+y^4(F+Gx^4)
    \end{split}
\end{equation}
The parameters $A$, $B$, and $C$ are set to be $\frac{1}{4}$, $\frac{2}{3}$, and $12$, respectively. The parameter $x_{\textrm{s}}$ and $V^{\ddagger}$ are set to be 0.2 and $\frac{3}{64000},$ respectively. The parameter $x_{\textrm{i}}$, that defines the coordinate of valley-ridge inflection (VRI) point $(x_{\textrm{i}},0.0)$, are changed over a range from 0.2 to 0.9. The parameters $D$, $F$, $G$ can then be determined by setting the coordinates $(x_{\textrm{p}},\pm y_{\textrm{p}})$ of the product wells at $(1.5,\pm 1.5)$ with energy $-0.025$ Hartree:
\begin{equation}
    \begin{cases}
      V(x_{\textrm{p}},\pm y_{\textrm{p}}) = -0.025\\
      \frac{dV(x,y)}{dx}|_{x=x_{\textrm{p}},y=\pm y_{\textrm{p}}} = 0\\
      \frac{dV(x,y)}{dy}|_{x=x_{\textrm{p}},y=\pm y_{\textrm{p}}} = 0
    \end{cases}       
\end{equation}
The values of $D$, $F$, $G$ are given in \cref{table:DFG} for all the different values of $x_{\textrm{i}}$.
\begin{table} [H]
    \centering
        \caption{Values for Parameters of D, F, G\label{table:DFG}}
        \begin{tabular} {|c|c|c|c|}
        \hline
        \hfil $x_{\textrm{i}}$ & \hfil $D$ & \hfil $F$ & \hfil $G$  \\ 
        \hline
        \hfil 0.2 & \hfil  0.01180 &  0.00369\hfil &-0.00006 \hfil  \\ 
        \hline
        \hfil 0.3 & \hfil 0.01278 & \hfil 0.00352 & \hfil -0.00002  \\
        \hline
        \hfil 0.4 & \hfil 0.01394 & \hfil 0.00333 & \hfil 0.00002  \\ 
        \hline
        \hfil 0.5 & \hfil 0.01534 & \hfil 0.00310 & \hfil 0.00006  \\ 
        \hline
        \hfil 0.6 & \hfil 0.01704 & \hfil 0.00281 & \hfil 0.00012 \\ 
        \hline
        \hfil 0.7 & \hfil  0.01917 & \hfil 0.00246 & \hfil 0.00019 \\ 
        \hline
        \hfil 0.8 & \hfil  0.02191 & \hfil 0.00200 & \hfil 0.00028  \\ 
        \hline
        \hfil 0.9 & \hfil 0.02556 & \hfil 0.00139 & \hfil 0.00040  \\ 
        \hline
        \end{tabular}
    \end{table}

\section{Langevin Dynamics Simulations}
\setcounter{page}{1}
\renewcommand{\thepage}{S\arabic{page}}

\subsection{Verlet-type Propagator with Magnetic Field}
Langevin dynamics were conducted with a Verlet-type propagator from Ref. \cite{Jensen2013} and we adapted it here to include the presence of a magnetic field through a second-order propagator. We will now recapitulate the algorithm. Consider the following Langevin equations,
\begin{equation}
\label{eqn:langevin_x}
    \begin{split}
        \dot{x} &= v_{\textrm{x}} \\
        \dot{y} &= v_{\textrm{y}} 
    \end{split}
\end{equation}
\begin{equation}
\label{eqn:langevin_v}
    \begin{split}
        \dot{v}_{\textrm{x}} &= -\gamma v_{\textrm{x}}-\frac{\nabla_{x}V(x,y)}{m} + \frac{\Omega}{m} v_{\textrm{y}} 
         + \frac{\eta_{\textrm{x}}(t)}{m} \\
        \dot{v}_{\textrm{y}} &= -\gamma v_{\textrm{y}} -\frac{\nabla_{y}V(x,y)}{m} - \frac{\Omega}{m} v_{\textrm{x}} 
         + \frac{\eta_{\textrm{y}}(t)}{m}
    \end{split}
\end{equation}
We begin by integrating \cref{eqn:langevin_v} over a small time interval $dt$ between two times $t_{n}$ and $t_{n+1}=t_{n}+dt$:
\begin{equation}
\begin{split}
      \int^{t_{n+1}}_{t_n}\dot{v}_{\textrm{x}} dt' &= \int^{t_{n+1}}_{t_n} \Big(-\frac{\nabla_{x}V(x,y)}{m} + \frac{\Omega}{m} v_{\textrm{y}} - \gamma v_{\textrm{x}} + \frac{\eta_{\textrm{x}}(t)}{m} \Big) dt' \\
      \int^{t_{n+1}}_{t_n}\dot{v}_{\textrm{y}}dt' & =  \int^{t_{n+1}}_{t_n} \Big(-\frac{\nabla_{y}V(x,y)}{m} -\frac{\Omega}{m} v_{\textrm{x}} - \gamma v_{\textrm{y}} + \frac{\eta_{\textrm{y}}(t)}{m}  \Big) dt'
\end{split}
\end{equation}
Without approximation, we can write 
\begin{equation}
    \label{eqn:vn+1}
    \begin{split}
        v^{n+1}_{\textrm{x}} - v^{n}_{\textrm{x}} &= - \gamma(x^{n+1}-x^{n})+ \frac{\Omega}{m} (y^{n+1}-y^{n})+\int^{t_{n+1}}_{t_n} -\frac{\nabla_{x}V(x,y)}{m}dt' + \int^{t_{n+1}}_{t_n} \frac{\eta_{\textrm{x}}(t)}{m}dt' \\
        v^{n+1}_{\textrm{y}} - v^{n}_{\textrm{y}} &= - \gamma(y^{n+1}-y^{n})- \frac{\Omega}{m} (x^{n+1}-x^{n})+\int^{t_{n+1}}_{t_n} -\frac{\nabla_{y}V(x,y)}{m}dt' + \int^{t_{n+1}}_{t_n} \frac{\eta_{\textrm{y}}(t)}{m}dt' 
    \end{split}
\end{equation}
Here, $\int^{t_{n+1}}_{t_n} \frac{\eta_{\textrm{x}}(t)}{m}dt'$ and  $\int^{t_{n+1}}_{t_n} \frac{\eta_{\textrm{y}}(t)}{m}dt'$ are two random numbers from two Gaussian distributions with mean of zero and standard deviation $\sqrt{\frac{2\gamma k_{\textrm{B}}T dt}{m}}$.
Integration of \cref{eqn:langevin_x} can be approximated with 
\begin{equation}
    \label{eqn:xn+1}
    \begin{split} 
    x^{n+1} - x^{n} \approx \frac{dt}{2} (v^{n+1}_{\textrm{x}}+v^{n}_{\textrm{x}}) \\
    y^{n+1} - y^{n} \approx \frac{dt}{2} (v^{n+1}_{\textrm{y}}+v^{n}_{\textrm{y}})
    \end{split}
\end{equation}
We make a similar approximation,
\begin{equation}
    \label{eqn:acn+1}
    \begin{split} 
    \int^{t_{n+1}}_{t_n} -\frac{\nabla_{x}V(x,y)}{m}dt' \approx \frac{dt}{2} (-\frac{\nabla_{x}V^{n+1}(x,y)}{m}-\frac{\nabla_{x}V^{n}(x,y)}{m}) \\
   \int^{t_{n+1}}_{t_n} -\frac{\nabla_{x}V(x,y)}{m}dt'\approx \frac{dt}{2} (-\frac{\nabla_{y}V^{n+1}(x,y)}{m}-\frac{\nabla_{y}V^{n}(x,y)}{m})
    \end{split}
\end{equation}
If we plug \cref{eqn:xn+1} and \cref{eqn:acn+1} into \cref{eqn:vn+1}, we find
\begin{equation}
    \label{eqn:vxn+1}
    \begin{split} 
    v^{n+1}_{\textrm{x}} &= v^{n}_{\textrm{x}} - \gamma(x^{n+1}-x^{n})+ \frac{\Omega dt}{2m} (v^{n+1}_{\textrm{y}}+v^{n}_{\textrm{y}})+\frac{dt}{2} (-\frac{\nabla_{x}V^{n+1}(x,y)}{m}-\frac{\nabla_{x}V^{n}(x,y)}{m})  +  \int^{t_{n+1}}_{t_n} \frac{\eta_{\textrm{x}}(t)}{m}dt' 
        \end{split}
\end{equation}
\begin{equation}
    \label{eqn:vyn+1}
    \begin{split} 
    v^{n+1}_{\textrm{y}} &= v^{n}_{\textrm{y}} - \gamma(y^{n+1}-y^{n})- \frac{\Omega dt}{2m} (v^{n+1}_{\textrm{x}}+v^{n}_{\textrm{x}})+\frac{dt}{2} (-\frac{\nabla_{y}V^{n+1}(x,y)}{m}-\frac{\nabla_{y}V^{n}(x,y)}{m})  +  \int^{t_{n+1}}_{t_n} \frac{\eta_{\textrm{y}}(t)}{m}dt'
    \end{split}
\end{equation}
Note that \cref{eqn:vxn+1} shows that $v_{\textrm{x}}^{n+1}$ depends on $v_{\textrm{y}}^{n+1}$ and \cref{eqn:vyn+1} shows that $v_{\textrm{y}}^{n+1}$ also depends on $v_{\textrm{x}}^{n+1}$. Hence, we can plug \cref{eqn:vyn+1} into \cref{eqn:vxn+1} and vice versa. After simplifying, we find:
\begin{equation}
    \label{eqn:final_v}
\begin{split}
    v^{n+1}_{\textrm{x}} =& \frac{1}{1+\frac{1}{4}(\frac{\Omega dt}{m})^2}\Bigg\{v^{n}_{\textrm{x}} - \gamma(x^{n+1}-x^{n}) +  \int^{t_{n+1}}_{t_n} \frac{\eta_{\textrm{x}}(t)}{m}dt' + \frac{dt}{2} \Big[-\frac{\nabla_{x}V^{n+1}(x,y)}{m}-\frac{\nabla_{x}V^{n}(x,y)}{m} \\
    & + \frac{\Omega}{m}2v^{n}_{\textrm{y}}-\frac{\Omega\gamma}{m}(y^{n+1}-y^{n})+\frac{\Omega}{m}\int^{t_{n+1}}_{t_n} \frac{\eta_{\textrm{y}}(t)}{m}dt'\Big]+ \frac{\Omega dt^2}{4m}\Big[-\frac{\Omega}{m}v^{n}_{\textrm{x}}  -\frac{\nabla_{y}V^{n+1}(x,y)}{m}-\frac{\nabla_{y}V^{n}(x,y)}{m}\Big]\Bigg\} 
\\
    v^{n+1}_{\textrm{y}} =& \frac{1}{1+\frac{1}{4}(\frac{\Omega dt}{m})^2}\Bigg\{v^{n}_{\textrm{y}} - \gamma(y^{n+1}-y^{n}) +  \int^{t_{n+1}}_{t_n} \frac{\eta_{\textrm{y}}(t)}{m}dt' + \frac{dt}{2} \Big[-\frac{\nabla_{y}V^{n+1}(x,y)}{m}-\frac{\nabla_{y}V^{n}(x,y)}{m} \\
    & - \frac{\Omega}{m}2v^{n}_{\textrm{x}}+\frac{\Omega\gamma}{m}(x^{n+1}-x^{n})-\frac{\Omega}{m}\int^{t_{n+1}}_{t_n} \frac{\eta_{\textrm{x}}(t)}{m}dt'\Big]- \frac{\Omega dt^2}{4m}\Big[\frac{\Omega}{m}v^{n}_{\textrm{y}}  -\frac{\nabla_{x}V^{n+1}(x,y)}{m}-\frac{\nabla_{x}V^{n}(x,y)}{m}\Big]\Bigg\} 
\end{split}
\end{equation}
Now to find $x^{n+1}$ and $y^{n+1}$, we plug \cref{eqn:vn+1} into 
\cref{eqn:xn+1}. After simplifying, we find:
\begin{equation}
\label{eqn:final_x}
    \begin{split} 
    x^{n+1} 
    & = x^{n} + \frac{dt}{ 1+ \frac{\gamma dt}{2} } v^{n}_{\textrm{x}} + \frac{\Omega dt^2}{2m( 1+ \frac{\gamma dt}{2})}v^{n}_{\textrm{y}} -\frac{\nabla_{x}V(x,y)dt^2}{2m( 1+ \frac{\gamma dt}{2})} + \frac{dt}{2( 1+ \frac{\gamma dt}{2})}\int^{t_{n+1}}_{t_n}\frac{\eta_{\textrm{x}}}{m}dt' 
    \\
    y^{n+1} 
    & = y^{n} + \frac{dt}{ 1+ \frac{\gamma dt}{2} } v^{n}_{\textrm{y}} - \frac{\Omega dt^2}{2m( 1+ \frac{\gamma dt}{2})}v^{n}_{\textrm{x}} -\frac{\nabla_{y}V(x,y)dt^2}{2m( 1+ \frac{\gamma dt}{2})} + \frac{dt}{2( 1+ \frac{\gamma dt}{2})}\int^{t_{n+1}}_{t_n}\frac{\eta_{\textrm{y}}}{m}dt' 
    \end{split} 
\end{equation}
\cref{eqn:final_x} and \cref{eqn:final_v} are the working equations for a second-order Verlet-type propagator including a magnetic field. On the one hand, if $\Omega=0$, \cref{eqn:final_x} and \cref{eqn:final_v} reduce to the working equations (Eq.19 and Eq.20) in Ref.\cite{Jensen2013}, which are also shown in \cref{eqn:no_mag}. The choice of $\gamma$ (with a maximum of 0.015 a.u.) and $dt$ (with a maximum of 10 a.u.) for this work satisfy the usual Verlet stability condition $\gamma dt<2$.\cite{Jensen2013}
\begin{equation}
\label{eqn:no_mag}
    \begin{split} 
    x^{n+1}_{\Omega=0}
    =&  x^{n} + \frac{dt}{ 1+ \frac{\gamma dt}{2} } v^{n}_{\textrm{x}}  -\frac{\nabla_{x}V(x,y)dt^2}{2m( 1+ \frac{\gamma dt}{2})} + \frac{dt}{2( 1+ \frac{\gamma dt}{2})}\int^{t_{n+1}}_{t_n}\frac{\eta_{\textrm{x}}}{m}dt' 
    \\
    y^{n+1}_{\Omega=0} 
    = & y^{n} + \frac{dt}{ 1+ \frac{\gamma dt}{2} } v^{n}_{\textrm{y}}  -\frac{\nabla_{y}V(x,y)dt^2}{2m( 1+ \frac{\gamma dt}{2})} + \frac{dt}{2( 1+ \frac{\gamma dt}{2})}\int^{t_{n+1}}_{t_n}\frac{\eta_{\textrm{y}}}{m}dt' 
    \\
    v^{n+1}_{x,\Omega=0} =& v^{n}_{\textrm{x}} + \frac{dt}{2} \Big[-\frac{\nabla_{x}V^{n+1}(x,y)}{m}-\frac{\nabla_{x}V^{n}(x,y)}{m} \Big] - \gamma(x^{n+1}-x^{n}) +  \int^{t_{n+1}}_{t_n} \frac{\eta_{\textrm{x}}(t)}{m}dt' 
\\
    v^{n+1}_{y,\Omega=0} =& v^{n}_{\textrm{y}}+ \frac{dt}{2} \Big[-\frac{\nabla_{y}V^{n+1}(x,y)}{m}-\frac{\nabla_{y}V^{n}(x,y)}{m} \Big] - \gamma(y^{n+1}-y^{n}) +  \int^{t_{n+1}}_{t_n} \frac{\eta_{\textrm{y}}(t)}{m}dt' 
    \end{split} 
\end{equation}
On the other hand, if $\gamma=0$, \cref{eqn:final_x} and \cref{eqn:final_v} reeduce to the working equations for the second-order Verlet-type propagator (Eq.10, Eq.13 and Eq.14)  in Ref.\cite{Spreiter1999}, which are also shown in \cref{eqn:no_fric}. The maximum time step $dt=10$ is  small enough for correctly  following  circular motion ($\frac{\Omega dt}{m}\ll2\pi$)\cite{Spreiter1999}; the maximum value of $\Omega$ is 35 a.u. and the effective mass is 2000 a.u. in this work. 
\begin{equation}
\label{eqn:no_fric}
    \begin{split}
    x^{n+1}_{\gamma=0}
     = &x^{n} + v^{n}_{\textrm{x}}dt  +\frac{dt^2}{2}\Big(-\frac{\nabla_{x}V^{n}(x,y)}{m} +\frac{\Omega}{m} v^{n}_{\textrm{y}} \Big)
    \\
    y^{n+1}_{\gamma=0} 
    = & y^{n} + v^{n}_{\textrm{y}}dt  +\frac{dt^2}{2}\Big(-\frac{\nabla_{y}V^{n}(x,y)}{m} -\frac{\Omega}{m} v^{n}_{\textrm{x}} \Big)
    \\
    v^{n+1}_{x,\gamma=0} =
    & \frac{1}{1+\frac{1}{4}(\frac{\Omega dt}{m} )^2} \Bigg\{ v^{n}_{\textrm{x}} + \frac{d t}{2}\Big[-\frac{\nabla_{x}V^{n+1}(x,y)}{m}-\frac{\nabla_{x}V^{n}(x,y)}{m} +\frac{2\Omega}{m} v^{n}_{\textrm{y}}\Big]\\
    &+\frac{\Omega dt^2}{4m}\Big[-\frac{\nabla_{y}V^{n+1}(x,y)}{m}-\frac{\nabla_{y}V^{n}(x,y)}{m} -\frac{\Omega}{m} v^{n}_{\textrm{x}}\Big]\Bigg\}\\
    v^{n+1}_{y,\gamma=0} = & \frac{1}{1+\frac{1}{4}(\frac{\Omega dt}{m})^2} \Bigg\{ v^{n}_{\textrm{y}} + \frac{dt}{2}\Big[-\frac{\nabla_{y}V^{n+1}(x,y)}{m}-\frac{\nabla_{y}V^{n}(x,y)}{m} -\frac{2\Omega}{m} v^{n}_{\textrm{x}}\Big]\\
    & - \frac{\Omega dt^2}{4m}\Big[-\frac{\nabla_{x}V^{n+1}(x,y)}{m}-\frac{\nabla_{x}V^{n}(x,y)}{m} +\frac{\Omega}{m} v^{n}_{\textrm{y}}\Big]\Bigg\}\\
    \end{split}
\end{equation}

\subsection{Simulation Convergence}
20000 trajectories were propagated at 300K for each calculation of product selectivity. A time step of 0.242 fs and a mass of 2000 a.u. were used. For all trajectories, the starting position of the particle was at the bottom of the reactant well with zero velocity. It is assumed that once the particle falls into one of the product wells, there is no turning back. To enforce this assumption, the simulations were terminated when the particle gets to $3k_{\textrm{B}}T$ lower in energy than the TS2. The numbers of trajectories that end up in $\textrm{P}_1$ and $\textrm{P}_2$ are counted and the product selectivity can be calculated as $S(\Omega,x_{\textrm{i}},\gamma,T,m) =\frac{N_{\textrm{P}_{1}}-N_{\textrm{P}_{2}}}{N_{\textrm{total}}}*100\%$. Table \ref{table:conv_traj} demonstrates the convergence of the product selectivity for  different combinations of Berry curvature elements ($\Omega$), $x$ coordinate of the VRI point ($x_{\textrm{i}}$), and friction coefficients ($\gamma$). Note that we assume the selectivity with zero magnetic field is zero (they are very close to zero $<\sim1\% $) and adjusted selectivity with non-zero magnetic field accordingly.
 \begin{table} [H]
    \centering
        \caption{Convergence of Product Selectivity (\%) with Numbers of Trajectories\label{table:conv_traj}}
        \begin{tabular} {|c|c|c|c|c|c|}
        \hline
         & &  &\multicolumn{3}{c|}{Number of Trajectories} \\
        \cline{4-6}
        $\Omega$ (a.u.) & $x_{\textrm{i}}$ (a.u.) & $\gamma$ (a.u.) &  20000 & 40000 &  60000 \\ 
        \hline
        \hfil 1.0 & \hfil 0.2 & \hfil 0.01 & \hfil -8.01 & \hfil-8.15& -8.08 \hfil \\ 
        \hline
        \hfil 1.0 & \hfil 0.6 & \hfil 0.0005 & \hfil -43.0 & \hfil-42.8& \hfil-43.1 \\ 
        \hline
        \hfil 1.0 & \hfil 0.6 & \hfil 0.01 & \hfil -9.46 & \hfil-9.59& \hfil-9.63 \\ 
        \hline
        \hfil 1.0 & \hfil 0.6 & \hfil 0.015 & \hfil -6.66& \hfil-6.60& \hfil-6.47 \\ 
        \hline
        \hfil 1.0 & \hfil 0.9 & \hfil 0.0005 & \hfil -46.6 & \hfil-45.9& \hfil-45.2 \\ 
        \hline
        \hfil 1.0 & \hfil 0.9 & \hfil 0.01 & \hfil -9.32 & \hfil-9.27& \hfil-9.19\\ 
        \hline
        \hfil 15.0 & \hfil 0.2 & \hfil 0.01 & \hfil -84.8 & \hfil-85.1& \hfil-85.1 \\ 
        \hline
        \hfil 15.0 & \hfil 0.6 & \hfil 0.01 & \hfil -91.4 & \hfil-91.3& \hfil-91.3 \\ 
        \hline
        \hfil 15.0 & \hfil 0.9 & \hfil 0.01 & \hfil -88.6 & \hfil-88.6& \hfil-88.4 \\ 
        \hline
        \hfil 30.0 & \hfil 0.2 & \hfil 0.01 & \hfil -99.2 & \hfil-99.1& \hfil-99.1 \\ 
        \hline
        \hfil 30.0 & \hfil 0.6 & \hfil 0.01 & \hfil -100 & \hfil -100 & \hfil-99.9 \\ 
        \hline
        \hfil 30.0 & \hfil 0.9 & \hfil 0.01 & \hfil -99.4 & \hfil-99.4& \hfil-99.2 \\ 
        \hline
        \end{tabular}\\
    \end{table}
Table \ref{table:conv_step} further validates the time step size used (0.242 fs = 10 a.u.) for the Langevin trajectories.
 \begin{table} [H]
    \centering
        \caption{Convergence of Product Selectivity (\%) with Time Step Sizes\label{table:conv_step}}
        \begin{tabular} {|c|c|c|c|c|c|}
        \hline
         & &  &\multicolumn{3}{c|}{Time Step Size (a.u.)} \\
        \cline{4-6}
        $\Omega$ (a.u.) & $x_{\textrm{i}}$ (a.u.) & $\gamma$ (a.u.) &  10 & 5 &  3 \\ 
        \hline
        \hfil 1.0 & \hfil 0.2 & \hfil 0.01 & \hfil -8.01 & \hfil-8.55& -8.07 \hfil \\ 
        \hline
        \hfil 1.0 & \hfil 0.6 & \hfil 0.0005 & \hfil -43.0 & \hfil-43.3&  -41.8\hfil \\ 
        \hline
                \hfil 1.0 & \hfil 0.6 & \hfil 0.01 & \hfil -9.46 & \hfil-9.61& -10.01 \hfil \\ 
        \hline
                \hfil 1.0 & \hfil 0.6 & \hfil 0.015 & \hfil -6.66 & \hfil-6.56& -6.41\hfil \\ 
        \hline
                \hfil 1.0 & \hfil 0.9 & \hfil 0.0005 & \hfil -46.6 & \hfil-46.0 & -45.2\hfil \\ 
        \hline
                \hfil 1.0 & \hfil 0.9 & \hfil 0.01 & \hfil -9.32 & \hfil-9.08& -8.81\hfil \\ 
        \hline
                \hfil 15.0 & \hfil 0.2 & \hfil 0.01 & \hfil -84.8 & \hfil-83.7& -85.1\hfil \\ 
        \hline
                \hfil 15.0 & \hfil 0.6 & \hfil 0.01 & \hfil -91.4 & \hfil-90.7& -91.7\hfil \\ 
        \hline
                \hfil 15.0 & \hfil 0.9 & \hfil 0.01 & \hfil -88.6 & \hfil -88.6 & -88.4 \hfil \\ 
        \hline
                \hfil 30.0 & \hfil 0.2 & \hfil 0.01 & \hfil -99.2 & \hfil-99.0& -99.1\hfil \\ 
        \hline
                \hfil 30.0 & \hfil 0.6 & \hfil 0.01 & \hfil -100 & \hfil-99.3& -100\hfil \\ 
        \hline
                \hfil 30.0 & \hfil 0.9 & \hfil 0.01 & \hfil -99.4 & \hfil-99.7& -99.5\hfil \\ 
        \hline
        \end{tabular}
    \end{table}

\section{Solutions to Langevin Dynamics with the Simplified Analytical Model}
The analytical solution for the Langevin equations (with potential Eq.(4)) is  complicated. However, we can simplify the form of the potential in Eq.(4) to $
    V^\textrm{s}(x,y) = -\sigma_{\textrm{x}}x+\frac{k_{\textrm{y}}y^2}{2} $, as shown in Eq.(6) in the main manuscript.
This potential has a negative slope $\sigma_{\textrm{x}}$ in the $x$ direction to mimic the energy drop between  TS1 and the VRI point and a quadratic dependence on $y$ (because the potential should be a valley before the VRI point). Although the potential forms a ridge after the VRI point, it still closely resembles a valley when it is close to the VRI point. 

To explain the impact of the effective magnetic field on the product selectivity, we calculate the distance between the two distributions of $y$ values with and without magnetic field. Specifically, we used the Hellinger distance because its value is bounded between 0 and 1. On the one hand, a Hellinger distance of zero between the distributions with and without magnetic field implies the magnetic field has no impact on the selectivity, which also corresponds to zero selectivity. On the other hand, a Hellinger distance of one means the presence of magnetic field completely changes the position distribution, which can be understood as 100\% of selectivity. 

As we use random Gaussian noise for the Langevin equations, we get a Gaussian distribution of $y$ values characterized by the mean $\langle y(t)\rangle_{t\rightarrow\infty}$ and variance $\langle (y(t)-\langle y(t)\rangle)^2\rangle_{t\rightarrow\infty}$. With the simplified potential $V^{\textrm{s}}(x,y)$, we calculated the mean and variance of $y$ values with and without magnetic field in the over-damped limit. Then we can calculate the Hellinger distance between the Gaussian distributions of $y$ values with and without magnetic field, which is directly compared with the product selectivity from the actual simulations.  
\subsection{Mean and Variance of y in the Over-Damped Limit}
\subsubsection{Zero Magnetic Field}
In the limit of large $\gamma$, the velocities can be assumed to be always relaxed and hence the acceleration terms of $\ddot{x}$ and $\ddot{y}$ are ignored.
\begin{equation}
    \label{eqn:overdamp_nomag}
    \begin{split}
        \gamma \dot{x} - \frac{\sigma_{\textrm{x}}}{m} &=  \frac{\eta_{\textrm{x}}(t) }{m} \\
        \gamma \dot{y} + \frac{k_{\textrm{y}}}{m}y &=  \frac{\eta_{\textrm{y}}(t) }{m}
    \end{split}
\end{equation}
The solution for $y$ is 
\begin{equation}
    y(t)=  y_0e^{-\frac{ k_{\textrm{y}}}{\gamma m}t} + \int \frac{ \eta_{\textrm{y}}(t')}{\gamma m}e^{-\frac{ k_{\textrm{y}}}{\gamma m}(t-t')} dt'
\end{equation}
Averaging over the random force, we get the average and variance to be,
\begin{equation}
  \begin{split}
    \langle y(t)\rangle&=  y_0e^{-\frac{ k_{\textrm{y}}}{\gamma m}t} \\
\langle (y(t)-\langle y(t)\rangle)^2\rangle &=  \frac{ k_{\textrm{B}}T}{k_{\textrm{y}}} \Big(1-e^{-2\frac{ k_{\textrm{y}}}{\gamma m}t}\Big)
   \end{split}
\end{equation}
Taking the steady-state limit,
$\langle y(t)\rangle_{t\rightarrow\infty} = 0$ and $\langle (y(t)-\langle y(t)\rangle)^2\rangle_{t\rightarrow\infty} =\frac{ k_{\textrm{B}}T}{k_{\textrm{y}}}  $

\subsubsection{Effective Magnetic Field}
Suppose now the particle feels an internal magnetic field in the $xy$ plane with a Berry curvature $\Omega$. We can easily write down the Langevin equations in the overdamped limit

\begin{equation}
    \begin{split}
        \dot{x} &=  \frac{\Omega}{m\gamma} \dot{y} + \frac{\sigma_{\textrm{x}}}{\gamma m} +  \frac{\eta_{\textrm{x}}(t)}{\gamma m} \\
        \dot{y} &=  - \frac{\Omega}{m\gamma} \dot{x} - \frac{k_{\textrm{y}}y}{\gamma m} + \frac{\eta_{\textrm{y}}(t)}{\gamma m}
    \end{split}
\end{equation}
To solve for $y$, again we  eliminate any contribution from $x$ and recover
\begin{equation}
\dot{y} =  - \frac{m\gamma k_{\textrm{y}}}{(m^2\gamma^2+\Omega^2)} y+ \frac{m\gamma \eta_{\textrm{y}}(t)}{(m^2\gamma^2+\Omega^2)} - \frac{ \sigma_{\textrm{x}}\Omega}{(m^2\gamma^2+\Omega^2)} -  \frac{\eta_{\textrm{x}}(t)\Omega}{(m^2\gamma^2+\Omega^2)}
\end{equation}
The solution is 
\begin{equation}
y(t)=  y_0e^{- \frac{m\gamma k_{\textrm{y}}}{(m^2\gamma^2+\Omega^2)}t}- \frac{\sigma_{\textrm{x}}\Omega}{m\gamma k_{\textrm{y}}}\Big(1-e^{- \frac{m\gamma k_{\textrm{y}}}{(m^2\gamma^2+\Omega^2)}t}\Big) + \int \Big[\frac{m\gamma \eta_{\textrm{y}}(t')}{(m^2\gamma^2+\Omega^2)} -  \frac{\eta_{\textrm{x}}(t')\Omega}{(m^2\gamma^2+\Omega^2)} \Big]e^{- \frac{m\gamma k_{\textrm{y}}}{(m^2\gamma^2+\Omega^2)}(t-t')} dt'
\end{equation}
Averaging over the random force, we find
\begin{equation}
\langle y(t)\rangle=  y_0e^{- \frac{m\gamma k_{\textrm{y}}}{(m^2\gamma^2+\Omega^2)}t}- \frac{\sigma_{\textrm{x}}\Omega}{m\gamma k_{\textrm{y}}}\Big(1-e^{- \frac{m\gamma k_{\textrm{y}}}{(m^2\gamma^2+\Omega^2)}t}\Big) 
\end{equation}
In the steady-state limit,  $\langle y\rangle_{t\to\infty}= - \frac{\sigma_{\textrm{x}}\Omega}{m\gamma k_{\textrm{y}}}$ \\
We can also calculate the variance of $y(t)$ averaged over the random force,
\begin{equation}
    \begin{split}
        \langle(y(t)-\langle y(t)\rangle)^2\rangle  
        = & e^{- \frac{2m\gamma k_{\textrm{y}}}{(m^2\gamma^2+\Omega^2)}t}\int dt''\int dt'\frac{e^{\frac{m\gamma k_{\textrm{y}}}{(m^2\gamma^2+\Omega^2)}(t'+t'')}}{(m^2\gamma^2+\Omega^2)^2}\Big[m^2\gamma^2 \langle \eta_{\textrm{y}}(t')\eta_{\textrm{y}}(t'')\rangle+\Omega^2 \langle \eta_{\textrm{x}}(t')\eta_{\textrm{x}}(t'') \rangle\Big] \\
        = &
         e^{- \frac{2m\gamma k_{\textrm{y}}}{(m^2\gamma^2+\Omega^2)}t}\int dt'\frac{2\gamma m k_{\textrm{B}}Te^{\frac{2m\gamma k_{\textrm{y}}}{(m^2\gamma^2+\Omega^2)}t'}}{(m^2\gamma^2+\Omega^2)} \\
         = &  \frac{k_{\textrm{B}}T}{k_{\textrm{y}}}\Big(1-e^{- \frac{2m\gamma k_{\textrm{y}}}{(m^2\gamma^2+\Omega^2)}t}\Big)
    \end{split}
\end{equation}
The equation assumes the random force is a white noise obeying $\langle \eta_\alpha(t)\eta_\beta(t')\rangle =2\gamma k_{\textrm{B}}T m\delta_{\alpha\beta}\delta(t-t')$. The terms that depend linearly on $\langle \eta_{\textrm{x}}(t')\rangle$ or $\langle \eta_{\textrm{y}}(t')\rangle$ are ignored because averages of the random force are zero. Again, taking the steady-state limit, $\langle(y(t)-\langle y(t)\rangle)^2\rangle_{t\to\infty}=\frac{k_{\textrm{B}}T}{k_{\textrm{y}}}$

\subsection{The Hellinger Distance}
\paragraph{The Overdamped Limit} In the limit of large $\gamma$ and $t\to\infty$, based on the means and variances obtained in the previous sections, the corresponding Gaussian distributions can be written down when there is no magnetic field,
\begin{equation}
    f(y) = \sqrt{\frac{k_{\textrm{y}}}{2\pi k_{\textrm{B}}T}}e^{-\frac{k_{\textrm{y}}}{2k_{\textrm{B}}T}y^2}
\end{equation}
and with magnetic field,
\begin{equation}
    g(y) = \sqrt{\frac{k_{\textrm{y}}}{2\pi k_{\textrm{B}}T}}e^{-\frac{k_{\textrm{y}}}{2k_{\textrm{B}}T}(y+\frac{\sigma_{\textrm{x}}\Omega}{m\gamma k_{\textrm{y}}})^2}
\end{equation}
The Hellinger distance $H$ can be then computed as
\begin{equation}
\begin{split}
    H^2(f,g) & = \frac{1}{2}\int \Big(\sqrt{f(y)} - \sqrt{g(y)}\Big)^2 dx \\
    & = 1-e^{-\frac{\sigma_{\textrm{x}}^2\Omega^2}{8 k_{\textrm{B}}Tk_{\textrm{y}}m^2\gamma^2}}
\end{split}
\end{equation}
\section{Values of $\sigma_{\textrm{x}}$ and $k_{\textrm{y}}$}
The slope of the potential in the $x$ direction ($\sigma_{\textrm{x}}$) is obtained by the ratio of the energy difference over the change in $x$ coordinate between  TS1 and the VRI point. To determine the value of $k_{\textrm{y}}$ for each choice of the VRI location, the distance between  TS1 and the VRI point along the $x$ axis were divided equally into 400 grid points. At each grid point, a harmonic fit can be performed along the $y$ direction to estimate the value of $k_{\textrm{y}}$. 
The value of $k_{\textrm{y}}$ changes as a function of $x$ and the closer the VRI point is to the TS1, the smaller the fitted $k_{\textrm{y}}$ value is. 
Thus, it is not straightforward to choose a single $k_{\textrm{y}}$. In this work, we used the $351^{th}$ grid point position to calculate $k_{\textrm{y}}$ for all VRI locations as this choice gave the lowest sum of the mean unsigned errors as compared to the simulation results. In other words, we select $k_{\textrm{y}}$ at a position $ \approx 87.8\%$ of the way from TS1 to the VRI point.  The specific values of $\sigma_{\textrm{x}}$ and $k_{\textrm{y}}$ for each VRI location are given in Table \ref{table:slope}.
 \begin{table} [H]
    \centering
        \caption{Values of $\sigma_{\textrm{x}}$ and $k_{\textrm{y}}$\label{table:slope}}
        \begin{tabular} {|c|c|c|}
        \hline
          $x_{\textrm{i}}$ & $\sigma_{\textrm{x}}$ &  $k_{\textrm{y}}$ \\ 
        \hline
        0.2 &  0.0029 & 0.00059 \\ 
        \hline
        0.3 &  0.0044 & 0.00094 \\ 
        \hline
        0.4 &  0.0058 & 0.00135 \\ 
        \hline
        0.5 &  0.0071 & 0.00185 \\ 
        \hline
        0.6 &  0.0083 & 0.00246 \\ 
        \hline
        0.7 &  0.0092 & 0.00323 \\ 
        \hline
        0.8 &  0.0100 & 0.00422 \\ 
        \hline
        0.9 &  0.0105 & 0.00554 \\ 
        \hline
        \end{tabular}
    \end{table}

\section{Computation of The Berry Curvature Within a  Generalized Hartree-Fock With Spin-Orbit Coupling Framework}
\subsection{Theory}
In this section, we give details as to how we evaluate the Berry curvature.
We start by writing down the electronic Hamiltonian of a molecular system with $N_{\mathrm{n}}$ number of nuclei and $N_{\mathrm{e}}$ number of electrons and also consider spin-orbit coupling (SOC),
\begin{equation}
    H_{\mathrm{e}} = V_{\mathrm{nn}}+T_{\mathrm{e}}+ V_{\mathrm{ne}} + V_{\mathrm{ee}} + V_{\mathrm{so}}
\end{equation}
The operator $V_{\mathrm{nn}}=\sum_{I>J}^{N_{\mathrm{n}}} \frac{Z_I Z_J}{\left|\mathbf{R}_I-\mathbf{R}_J\right|}$ accounts for the repulsion between nucleus $I$ and nucleus $J$ with atomic number $Z_{I}$ and $Z_{J}$ , and nuclear coordinates $\mathbf{R}_I$ and $\mathbf{R}_J$, respectively.  The operator $T_{\mathrm{e}} = -\frac{1}{2} \sum_i^{N_{\mathrm{e}}} \nabla_i^2$ is the kinetic energy operator for electron $i$. The operator $V_{\mathrm{ne}} =-\sum_i^{N_{\mathrm{e}}} \sum_I^{N_{\mathrm{n}}} \frac{Z_I}{\left|\mathbf{r}_i-\mathbf{R}_I\right|} $ captures the Coulomb attraction between electrons with electronic coordinate $\mathbf{r}_i$ and nuclei. The operator $V_{\mathrm{ee}} = \sum_{i>j}^{N_{\mathrm{e}}} \frac{1}{\left|\mathbf{r}_i-\mathbf{r}_j\right|} $ captures the electron-electron interaction. For the spin-orbit coupling (SOC) operator $V_{\mathrm{so}}$, we use only the one-electron component of the Breit-Pauli form $V_{\mathrm{so}}=-\frac{\alpha_0^2}{2} \sum_{i,I} \frac{\mathrm{Z}_{I}}{\left|\mathbf{r}_{iI}\right|^3}\left(\mathbf{r}_{iI} \times \mathbf{p}_i\right) \cdot \mathbf{s}_i$,\cite{Abegg1975} where $\alpha_0$ is the fine structure constant. The variable $\mathbf{r}_{iI}$ is the distance between the electron $i$ and nucleus $I$, and $\mathbf{s}_i$ and $\mathbf{p}_i$ is the spin operator and momentum of electron $i$, respectively.

Under the generalized Hartree-Fock framework (GHF),\cite{Pople1977,Fukutome1981,Goings2015} the electronic wavefunction ansatz $|\Psi\rangle$ is a single slater determinant made up of spinor orbitals $\Phi_i (\mathbf{r})$, each of which can expanded into a linear combination of atomic orbitals $\phi_{\mu}$,
\begin{equation}
    \Phi_i (\mathbf{r}) = \sum_{\mu\tau} c^{\tau}_{\mu}\phi_{\mu}(\mathbf{r}) .
\end{equation}
From this point onwards, the indices $ijkl$, $\mu\nu\sigma\lambda$, and $\tau\kappa\eta\xi$ will be used for spinor orbitals, atomic orbitals, and spins, respectively. Within this wavefunction ansatz, the total GHF+SOC energy expression can be written as, 
\begin{equation}
\label{eqn:E}
E^{\textrm{GHF+SOC}} = \sum_{\mu\nu\tau\kappa}h_{\mu\nu} P^{\kappa\tau}_{\nu\mu} \delta_{\tau\kappa} + \sum_{\mu\nu\tau\kappa}h^{so,\tau\kappa}_{\mu\nu} P^{\kappa\tau}_{\nu\mu} + \frac{1}{2}\sum_{\mu\nu\sigma\lambda\tau\kappa} G^{\tau\kappa}_{\mu\nu} P^{\kappa\tau}_{\nu\mu}
\end{equation}
where the density matrix $P^{\tau\kappa}_{\mu\nu} = \sum_{i} c^{\tau}_{\mu i}c^{\kappa*}_{\nu i}$. $h_{\mu\nu}$ is the conventional GHF one-electron term. $h^{\textrm{so},\tau\kappa}_{\mu\nu}$ is the one-electron SOC term and can be written explicitly in the matrix form,
\begin{equation}
    h^{\textrm{so}} = \begin{pmatrix}
    h^{\textrm{so},\alpha\alpha}_{\mu\nu} &  h^{\textrm{so},\alpha\beta}_{\mu\nu} \\
    h^{\textrm{so},\beta\alpha}_{\mu\nu} &  h^{\textrm{so},\beta\beta}_{\mu\nu} \\ 
   \end{pmatrix} 
   = -\frac{\alpha_{0}^2}{4}\begin{pmatrix}
    L^{z}_{\mu\nu} &  L^{x}_{\mu\nu} - iL^{y}_{\mu\nu} \\
    L^{x}_{\mu\nu}+iL^{y}_{\mu\nu} &  -L^{z}_{\mu\nu} \\ 
\end{pmatrix}
\end{equation}
where the specific expressions for the $L_{\mu\nu}$ integrals can be found in Ref\cite{King1988,Bellonzi2019}. The 2-electron component in \cref{eqn:E} is given as $G^{\tau\kappa}_{\mu\nu} = \sum_{\sigma\lambda\eta\xi}P^{\eta\xi}_{\sigma\lambda}\Big[(\mu\nu|\lambda\sigma)\delta_{\tau\kappa} \delta_{\xi\eta} - (\mu\sigma|\lambda     \nu)\delta_{\kappa\xi} \delta_{\eta\tau}\Big]$.

According to Eq.(19) and Eq.(61) from Ref.\cite{Culpitt2022}, the expression for the Berry curvature elements within a complex GHF framework is
\begin{equation}
\label{eqn:bc}
\begin{split}
    \Omega_{I \alpha J \beta} &=-2\hbar \mathrm{Im}\langle\nabla_{I \alpha} \Psi |\nabla_{J \beta} \Psi \rangle \\
=&-2 \hbar \operatorname{Im}\left[\sum_{i}\left\langle\Phi_{i}^{(I \alpha)} \mid \Phi_{i}^{(J \beta)}\right\rangle+\sum_{i a}\left\langle\Phi_{a} \mid \Phi_{i}^{(J \beta)}\right\rangle U_{a i}^{I \alpha *}\right.\\
&+\sum_{i a}\left\langle\Phi_{i}^{(I \alpha)} \mid \Phi_{a}\right\rangle U_{a i}^{J \beta}+\sum_{i a} U_{a i}^{I \alpha *} U_{a i}^{J \beta} \left.-\sum_{i j}\left\langle\Phi_{i}^{(I \alpha)} \mid \Phi_{j}\right\rangle\left\langle\Phi_{j} \mid \Phi_{i}^{(J \beta)}\right\rangle\right]
\end{split}
\end{equation}
where the notation $\Phi_{i}^{(I \alpha)}$ indicates that we are taking the derivative of the spinor orbital $\Phi_{i}$ with respect to the $\alpha$ coordinate of nucleus $I$, while holding the orbital coefficients unchanged. Here, the indices $\alpha$ and $\beta$ represent  Cartesian directions $\{x,y,z\}$. $U^{I\alpha}_{ni}$ is defined as $\frac{\partial c_{\mu i}^\tau}{\partial R_{I\alpha}}=c_{\mu i}^{\tau, I\alpha}=\sum_n c_{\mu n}^\tau U_{n i}^{I\alpha}$, where the sum over $n$  includes all spinor orbitals; this quantity is calculated by solving the couple-perturbed SCF (CPSCF) equations.\cite{Culpitt2022} 

\subsection{Implementation Details}
We have implemented GHF+SOC, the corresponding CPSCF, and the orbital Hessian for stability analysis in a local branch of Q-Chem 6.0.\cite{Epifanovsky2021} For all of these functionalities, we used complex orbital coefficients $c^{\tau}_{\mu n}$ and real atomic basis functions $\phi_{\mu}(\mathbf{r})$. All the matrix elements that go into \cref{eqn:bc} are computed analytically, except the integral $\left\langle\Phi_{i}^{(I \alpha)} \mid \Phi_{i}^{(J \beta)}\right\rangle$ which at present is computed by finite difference with a five-point stencil. We have verified that our Berry curvature as calculated analytically agree with our finite difference calculations. The specific equations for evaluating the Berry curvature by finite difference are given in Eq.(85) and Eq.(86) Ref.\cite{Culpitt2021}.  Details of this Berry curvature calculation will be reported in a future publication. 

\subsection{Methoxy Radical Isomerization}
\subsubsection{Cartesian Coordinates of Relevant Geometries}
The geometries at stationary and the valley-ridge inflection (VRI) points on the potential energy surface of methoxy radical isomerization were obtained from previously\cite{Taketsugu1996}  optimized unrestricted Hartree-Fock (UHF) with 6-31G(d,p) basis set. To estimate the reaction path between transition state 1 (TS1) and transition state 2 (TS2), we interpolated the internal coordinates between TS1 and TS2 and obtained 10 geometries in between. The Cartesian coordinates for the reactant (R), TS1, TS2, VRI, products (P1 and P2), and the interpolated geometries are given below with unit of Angstrom.
\begin{enumerate}
\item \textbf{R} \\
        C        -0.0000839903    0.0000000000    0.0001450441 \\
        O        -0.0000828687    0.0000000000    1.3825167119 \\
        H         1.0456407148    0.0000000000   -0.3023634452 \\
        H        -0.4717516296   -0.8921927701   -0.4016919625 \\
        H        -0.4717516296    0.8921927701   -0.4016919625
\item \textbf{TS1} \\
          C       0.0000000000     0.0000000000     0.0000000000 \\
          O       0.0000000000     0.0000000000     1.3670000000 \\
          H       1.0154851183     0.0000000000     0.7543308123\\
          H      -0.2481603576    -0.9261470631    -0.4948828693 \\ 
          H      -0.2481603576     0.9261470631    -0.4948828693
\item \textbf{TS2} \\
          C       0.0000000000     0.0000000000     0.0000000000 \\
          O       0.0000000000     0.0000000000     1.3651050000 \\
          H       0.8836643646     0.0000000000     1.6971035075 \\
          H      -0.2499942581    -0.9302257507    -0.4817441851 \\
          H      -0.2499942581     0.9302257507    -0.4817441851
\item \textbf{P1}  \\
          O       0.0000000000     0.0000000000     0.0000000000 \\
          H       0.0000000000     0.0000000000     0.9423910000 \\
          C       1.2720549702     0.0000000000    -0.4739351520 \\
          H       1.3199540237    0.0309204231    -1.5460012749 \\
          H       2.0216170261    0.5247607414     0.0964007436
\item \textbf{P2}  \\
          O       0.0000000000     0.0000000000     0.0000000000 \\
          H       0.0000000000     0.0000000000     0.9423910000 \\
          C       1.2720549702     0.0000000000    -0.4739351520 \\
          H       1.3199540237     -0.0309204231    -1.5460012749 \\
          H       2.0216170261     -0.5247607414     0.0964007436
\item \textbf{VRI} \\
          C       0.0000000000     0.0000000000     0.0000000000\\
          O       0.0000000000     0.0000000000     1.3720000000\\
          H       0.9542514219     0.0000000000     1.0797616514\\
          H      -0.2250338281    -0.9301783583    -0.4918373695\\
          H      -0.2250338281     0.9301783583    -0.4918373695
          
\item \textbf{Interpolated Step 1}  \\
          C       0.0000000000     0.0000000000     0.0000000000 \\
          O       0.0000000000     0.0000000000     1.3668277273 \\
          H       1.0294861763     0.0000000000     0.8324571851 \\
          H      -0.2483287318    -0.9265250784    -0.4936898329 \\
          H      -0.2483287318     0.9265250784    -0.4936898329
\item \textbf{Interpolated Step 2}  \\
          C       0.0000000000     0.0000000000     0.0000000000 \\
          O       0.0000000000     0.0000000000     1.3666554546 \\
          H       1.0387065820     0.0000000000     0.9129435062 \\
          H      -0.2484967757    -0.9269016486    -0.4924965133 \\
          H      -0.2484967757     0.9269016486    -0.4924965133
\item \textbf{Interpolated Step 3} \\ 
          C       0.0000000000     0.0000000000     0.0000000000 \\
          O       0.0000000000     0.0000000000     1.3664831818 \\
          H       1.0429787670     0.0000000000     0.9955258218 \\
          H      -0.2486644890    -0.9272767738    -0.4913029120 \\
          H      -0.2486644890     0.9272767738    -0.4913029120
\item \textbf{Interpolated Step 4} \\
          C       0.0000000000     0.0000000000     0.0000000000 \\
          O       0.0000000000     0.0000000000     1.3663109091 \\
          H       1.0421489024     0.0000000000     1.0799305078 \\
          H      -0.2488318715    -0.9276504539    -0.4901090310 \\
          H      -0.2488318715     0.9276504539    -0.4901090310
\item \textbf{Interpolated Step 5}  \\
          C       0.0000000000     0.0000000000     0.0000000000 \\
          O       0.0000000000     0.0000000000     1.3661386364 \\
          H       1.0360774051     0.0000000000     1.1658749578 \\
           H      -0.2489989228    -0.9280226889    -0.4889148722 \\
          H      -0.2489989228     0.9280226889    -0.4889148722
\item \textbf{Interpolated Step 6}  \\
          C       0.0000000000     0.0000000000     0.0000000000 \\
          O       0.0000000000     0.0000000000     1.3659663636 \\
          H       1.0246394119     0.0000000000     1.2530682971 \\
          H      -0.2491656426    -0.9283934788    -0.4877204373 \\
          H      -0.2491656426     0.9283934788    -0.4877204373
\item \textbf{Interpolated Step 7} \\
          C       0.0000000000     0.0000000000     0.0000000000 \\
          O       0.0000000000     0.0000000000     1.3657940909 \\
          H       1.0077252175     0.0000000000     1.3412121187 \\
          H      -0.2493320306    -0.9287628236    -0.4865257281 \\
          H      -0.2493320306     0.9287628236    -0.4865257281
\item \textbf{Interpolated Step 8} \\ 
          C       0.0000000000     0.0000000000     0.0000000000 \\
          O       0.0000000000     0.0000000000     1.3656218182 \\
          H       0.9852406773     0.0000000000     1.4300012423 \\
          H      -0.2494980865    -0.9291307232    -0.4853307465 \\
          H      -0.2494980865     0.9291307232    -0.4853307465
\item \textbf{Interpolated Step 9} \\ 
          C       0.0000000000     0.0000000000     0.0000000000 \\
          O       0.0000000000     0.0000000000     1.3654495454 \\
          H       0.9571075734     0.0000000000     1.5191244934 \\
          H      -0.2496638099    -0.9294971775    -0.4841354942 \\
          H      -0.2496638099     0.9294971775    -0.4841354942
\item \textbf{Interpolated Step 10} \\
          C       0.0000000000     0.0000000000     0.0000000000 \\
          O       0.0000000000     0.0000000000     1.3652772727 \\
          H       0.9232639425     0.0000000000     1.6082655003 \\
          H      -0.2498292005    -0.9298621867    -0.4829399732 \\
          H      -0.2498292005     0.9298621867    -0.4829399732
\end{enumerate}

\subsubsection{Complex GHF+SOC Results}
We computed the complex GHF+SOC energies with 6-31G(d,p) basis set for the geometries of interest given in S5.3.1. We confirmed each of the optimized SCF solution is a minimum by diagonalizing the complex GHF+SOC orbital Hessian. We provided the GHF+SOC energies, the SOC energy, the lowest three eigenvalue from the orbital Hessian, and $\langle S^2 \rangle$ in \cref{table:CGHF+SOC}. The GHF+SOC energy differs from UHF only on the order of $10^{-6}$ Hartree. The small eigenvalues suggest the SCF minima in the orbital space are very flat, possibly due to the relatively small magnitude of SOC effect. Note that without SOC, the first two eigenvalues from the orbital Hessian would be on the order of $10^{-12}$ or smaller. The $\langle S^2 \rangle$ values change approximately from 0.80 to 0.76 along the TS1-to-TS2 path, which is close to the $\langle S^2 \rangle$ value of 0.75 for a pure doublet. The time-reversed solution gives the same GHF+SOC energies but a Berry curvature matrix negative of the original matrix.

\begin{table} [H]
    \centering
        \caption{Complex GHF+SOC Results in Hartree\label{table:CGHF+SOC}}
        \begin{tabular} {|c|c|c|c|c|c|c|}
        \hline
        \hfil  & \hfil GHF+SOC Energies  & \hfil SOC Energy  & \hfil Eigenvalue 1  & \hfil Eigenvalue 2  & \hfil Eigenvalue 3 & \hfil $\langle S^2 \rangle$ \\ 
        \hline
        \hfil TS1 & \hfil  -114.3403272695   & \hfil $-9.72\mathrm{e}{-6}$ & \hfil $1.29\mathrm{e}{-7}$ & \hfil $3.54\mathrm{e}{-7}$& \hfil 0.121 & \hfil 0.796 \\
        \hline
        \hfil Step 1 & \hfil  -114.3420444373   & \hfil $-9.14\mathrm{e}{-6}$ & \hfil $7.46\mathrm{e}{-8}$& \hfil $2.53\mathrm{e}{-7}$ & \hfil 0.140 &\hfil 0.797\\
        \hline
        \hfil Step 2 & \hfil -114.3473033282  & \hfil $-8.67\mathrm{e}{-6}$ & \hfil $4.19\mathrm{e}{-8}$ & \hfil $1.76\mathrm{e}{-7}$& \hfil 0.159 &\hfil 0.790\\
        \hline
        \hfil Step 3 & \hfil -114.3556876981 & \hfil $-8.30\mathrm{e}{-6}$ & \hfil $2.68\mathrm{e}{-8}$ & \hfil $1.21\mathrm{e}{-7}$& \hfil 0.176&\hfil 0.778\\
        \hline
        \hfil Step 4 & \hfil -114.3661032327  & \hfil $-8.01\mathrm{e}{-6}$ & \hfil $2.31\mathrm{e}{-8}$ & \hfil $8.49\mathrm{e}{-8}$& \hfil 0.187 &\hfil 0.768\\
        \hline
        \hfil Step 5 & \hfil -114.3772322329  & \hfil $-7.78\mathrm{e}{-6}$ & \hfil $2.42\mathrm{e}{-8}$ & \hfil $6.36\mathrm{e}{-8}$ & \hfil 0.203 &\hfil 0.763\\
        \hline
        \hfil Step 6 & \hfil  -114.3879289229 & \hfil $-7.59\mathrm{e}{-6}$ & \hfil $2.63\mathrm{e}{-8}$ & \hfil $5.14\mathrm{e}{-8}$ & \hfil 0.223 &\hfil 0.761\\
        \hline
        \hfil Step 7 & \hfil  -114.3973571337  & \hfil  $-7.45\mathrm{e}{-6}$ & \hfil $2.81\mathrm{e}{-8}$&\hfil $4.44\mathrm{e}{-8}$ & \hfil 0.235 &\hfil 0.760\\
        \hline
        \hfil Step 8 & \hfil -114.4049671759 & \hfil  $-7.33\mathrm{e}{-6}$ & \hfil $2.93\mathrm{e}{-8}$ & \hfil $4.02\mathrm{e}{-8}$ & \hfil 0.239&\hfil 0.760\\ 
        \hline
        \hfil Step 9 & \hfil -114.4104465184 & \hfil  $-7.24\mathrm{e}{-6}$ & \hfil $3.02\mathrm{e}{-8}$& \hfil $3.79\mathrm{e}{-8}$ & \hfil 0.241 &\hfil 0.760\\ 
        \hline
        \hfil Step 10 & \hfil -114.4136839264 & \hfil  $-7.17\mathrm{e}{-6}$ & \hfil $3.10\mathrm{e}{-8}$& \hfil $3.69\mathrm{e}{-8}$ & \hfil 0.242 &\hfil 0.760\\ 
        \hline
        \hfil TS2 & \hfil -114.4147363353 & \hfil  $-7.13\mathrm{e}{-6}$ & \hfil $3.19\mathrm{e}{-8}$ & \hfil $3.67\mathrm{e}{-8}$ & \hfil 0.243 &\hfil 0.760\\ 
        \hline
        \hfil VRI & \hfil -114.3713220978  & \hfil  $-7.65\mathrm{e}{-6}$ & \hfil $3.14\mathrm{e}{-8}$ & \hfil $7.11\mathrm{e}{-8}$& \hfil 0.213 &\hfil 0.764\\ 
        \hline
        \end{tabular}
    \end{table}

The 2-induced norm of Berry curvature matrix $\vert\vert\mathbf{\Omega}\vert\vert_{2}$ computed with numerical procedure and the analytic procedure (\cref{eqn:bc}) is given in \cref{table:BC}. The analytic implementation of Berry curvature gives results that match the numerical results, which confirmed the correct analytic implementation. 
\begin{table} [H]
    \centering
        \caption{Comparison between Analytic and Numerical Computation of $\vert\vert\mathbf{\Omega}\vert\vert_{2}$ \label{table:BC}}
        \begin{tabular} {|c|c|c|}
        \hline
        \hfil  & \hfil Analytic $\vert\vert\mathbf{\Omega}\vert\vert_{2}$ &  \hfil Numerical $\vert\vert\mathbf{\Omega}\vert\vert_{2}$  \\ 
        \hline
        \hfil TS1 & \hfil  0.1223    & \hfil  0.1223\\ 
        \hline
        \hfil Step 1 & \hfil  0.1546  & \hfil  0.1546 \\ 
        \hline
        \hfil Step 2 & \hfil 0.2165 & \hfil 0.2164  \\
        \hline
        \hfil Step 3 & \hfil 0.2699 & \hfil 0.2699  \\ 
        \hline
        \hfil Step 4 & \hfil 0.2498  & \hfil 0.2498 \\ 
        \hline
        \hfil Step 5 & \hfil 0.2025  & \hfil 0.2025 \\ 
        \hline
        \hfil Step 6 & \hfil  0.1763 & \hfil  0.1764\\ 
        \hline
        \hfil Step 7 & \hfil  0.1679 & \hfil  0.1679 \\ 
        \hline
        \hfil Step 8 & \hfil 0.1656 & \hfil 0.1655\\ 
        \hline
        \hfil Step 9 & \hfil 0.1624  & \hfil 0.1624 \\ 
        \hline
        \hfil Step 10 & \hfil 0.1553 & \hfil 0.1552  \\ 
        \hline
        \hfil TS2 & \hfil 0.1432 & \hfil 0.1430  \\ 
        \hline
        \hfil VRI & \hfil 0.1684  & \hfil 0.1683  \\ 
        \hline
        \end{tabular}
    \end{table}
\subsubsection{TS1-to-TS2 and TS2-to-P1 Directions}
To estimate the unit vectors pointing from TS1 to TS2, we performed the normal mode analysis at TS1 and used the normal mode corresponding to the imaginary frequency, which is given in Fig. S1(a). To estimate the unit vectors pointing from TS2 to P1, we performed the normal mode analysis at TS2 and used the normal mode corresponding to the imaginary frequency computed at TS2, which is given in Fig. S1(b). 

\begin{center}
\includegraphics[width=0.58\columnwidth]{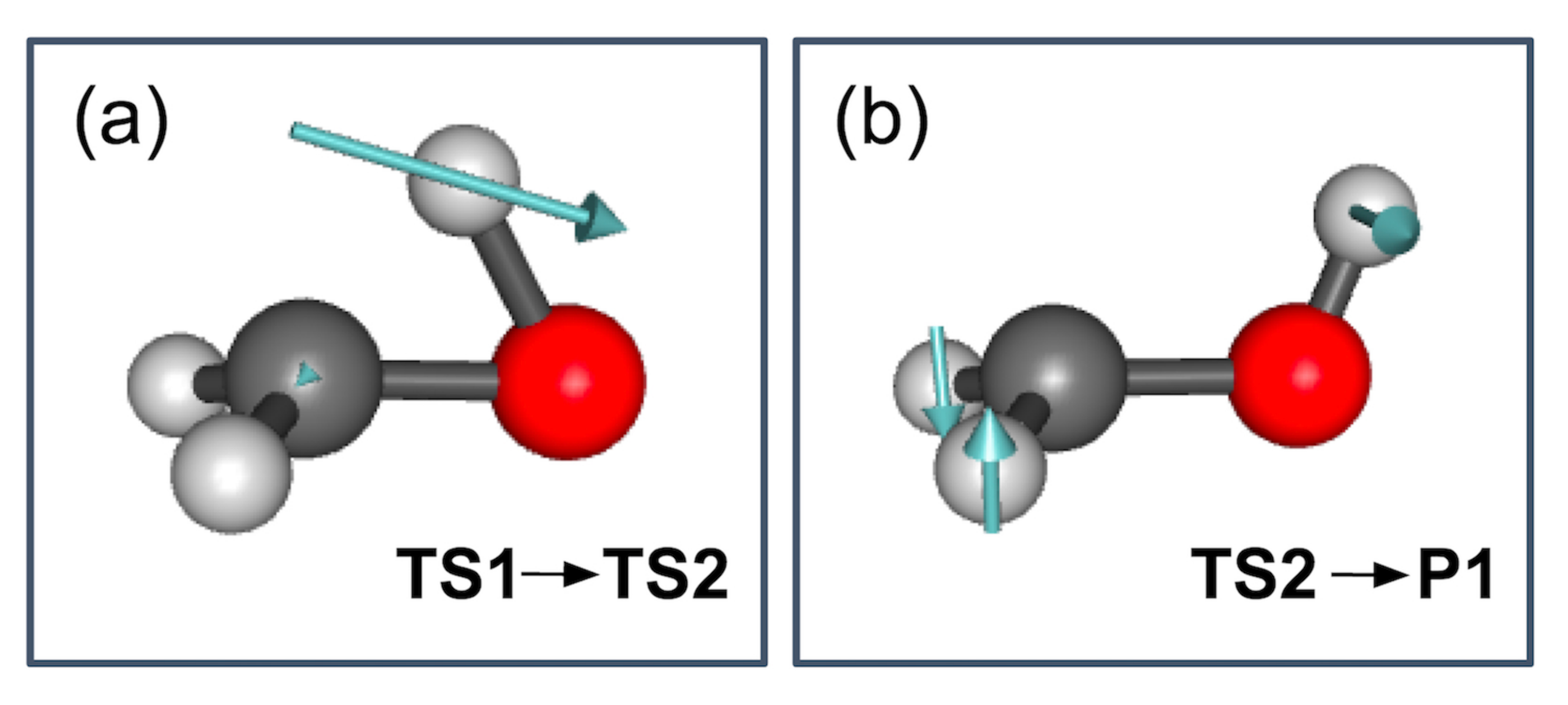} \\
\end{center}
Figure S1. (a) Normal mode corresponding to the imaginary frequency computed at TS1, which is used as the velocity vector for the Berry force computation (b) Normal mode corresponding to the imaginary frequency computed at TS2, which is used to approximate the unit vector pointing from TS2 to P1.

\bibliography{SI}